# A comparison of classical interatomic potentials applied to highly concentrated aqueous lithium chloride solutions

Ildikó Pethes

Wigner Research Centre for Physics, Hungarian Academy of Sciences, H-1525 Budapest, POB 49, Hungary

E-mail address: pethes.ildiko@wigner.mta.hu

Abstract

Aqueous lithium chloride solutions up to very high concentrations were investigated in classical molecular dynamics simulations. Various force fields based on the 12-6 Lennard-Jones model, parametrized for non-polarizable water solvent molecules (SPC/E, TIP4P, TIP4PEw), were inspected. Twenty-nine combinations of ion-water interaction models were examined at four different salt concentrations. Densities, static dielectric constants and self-diffusion coefficients were calculated. Results derived from the different force fields scatter over a wide range of values. Neutron and X-ray weighted structure factors were also calculated from the radial distribution functions and compared with experimental data. It was found that the agreement between calculated and experimental curves is rather poor for several investigated potential models, even though some of them have previously been applied in computer simulations.

None of the investigated models yield satisfactory results for all the tested quantities. Only two parameter sets provide acceptable predictions for the structure of highly concentrated aqueous LiCl solutions. Some approaches for adjusting potential parameters, such as those of Aragones [Aragones et al., J. Phys. Chem. B 118 (2014) 7680] and Pluharova [Pluharova et al, J. Phys. Chem. A 117 (2013) 11766], were tested as well; the simulations presented here underline their usefulness. These refining methods are suited to obtain more appropriate ion/water potentials.

# 1. Introduction

Aqueous electrolyte solutions receive enormous attention because of their great importance in physical chemistry, geochemistry as well as in environmental and industrial fields. They are essential elements for biochemical reactions in living organisms, thus to understand and predict their properties is fundamental for biochemical research. Biological processes mostly take place at low and moderate concentrations, industrial and geochemical applications require information about the more concentrated solutions up to the solubility limit. Consequently, simple electrolyte solutions as well as complex biomolecules are in the center of the interest of multiple publications (see e.g. [1-3] etc.).

Classical molecular dynamics simulation is nowadays one of the (perhaps the) most popular tools in the study of these systems. Several interaction models (force field, FF) have been developed in the past 30 years to describe simple aqueous solutions (see e. g. in Refs. [4-9]), they are also widely used for the study of more complex systems [10-13]. These interatomic potentials are suited for different water models, from the simplest 3-sites, rigid, non-polarizable models (such as SPC [14]) to the polarizable, four-site models (such as SWM4-NDP [15] with Drude oscillators, or BK3 [16] with a Gaussian charge distribution). The van der Waals interactions between atoms are usually taken into account by the 12-6 Lennard-Jones (LJ) potential, which requires 2 parameters for each type of atom pairs [17]. More complex models (such as 12-6-4 LJ model [18], the Buckingham (or EXP6) interaction model [19] or the polarizable ion FF models [20,21]) with more parameters are also used to describe ionic interactions.

Various ionic models are also routinely applied in the simulations of simple aqueous solutions or biomolecular systems. There are several comprehensive studies (such as [4,22,23]) concerning the adaptability of these models, yet the ionic parameter sets are often chosen almost randomly, and/or ionic force field parameters developed for other (different) water models are combined (see e.g. [24-27]).

In a recent paper [23] thirteen of the most common (simple and computationally low cost) 12-6 LJ ionic force fields together with the widely used SPC/E water model [28] have been investigated for aqueous NaCl (from dilute to concentrated) solutions. The authors showed that most of the examined force fields are unable to describe adequately even basic properties of NaCl solutions in the entire concentration range. In another study [29] the concentration and temperature dependence of the self-diffusion of water in nine different electrolyte solutions was investigated. They found that none of the seven combinations of ionic and water models employed can reproduce the experimentally observed concentration dependence.

It is not expected that a single model – especially a simple one, like the 12-6 LJ models – can predict all investigated physical and chemical properties well. But for each application it is important to make sure that the most proper one for that specific target has been chosen.

Aqueous LiCl solutions are among the most frequently investigated electrolyte solutions. LiCl has been considered for thermodynamic and structural studies due to its high solubility [30]. $Li^+$, which is the smallest cation, plays an important role in biological, medical and technical applications (see e.g. [31,32]). Several papers using MD simulations with different force field parameters have been published (e.g. [25-27, 33, 34-47]). However to the best of my knowledge, there is no study which collects and compares the available ionic force fields for aqueous LiCl solutions.

In this report 12-6 LJ ionic interatomic models of LiCl solutions, developed for rigid, non-polarizable water models, are investigated. They are used for simulating highly concentrated aqueous solutions. Their predictions about some basic physical and chemical properties (density, static dielectric constant, self-diffusion coefficients) are examined and compared to experimental values. Structure factors calculated from the atomic configurations are compared to total scattering structure factors from neutron and X-ray diffraction measurements reported earlier [48]. Some combinations of force field parameters applied in recent publications are also tested.

## 2. Methods

### 2.1 Molecular force fields

Pairwise additive non-polarizable intermolecular potentials were tested, which describe the interaction energy between two atoms or ions via the Coulomb potential

$$V_C(r_{ij}) = \frac{1}{4\pi\varepsilon_0} \frac{q_i q_j}{r_{ij}} \qquad (1)$$

and the 12-6 LJ potential:

$$V_{LJ}(r_{ij}) = 4\varepsilon_{ij}\left[\left(\frac{\sigma_{ij}}{r_{ij}}\right)^{12} - \left(\frac{\sigma_{ij}}{r_{ij}}\right)^{6}\right] = \varepsilon_{ij}\left[\left(\frac{R_{\min,ij}}{r_{ij}}\right)^{12} - 2\left(\frac{R_{\min,ij}}{r_{ij}}\right)^{6}\right] = \frac{C_{12,ij}}{r_{ij}^{12}} - \frac{C_{6,ij}}{r_{ij}^{12}} \qquad (2)$$

Here $r_{ij}$ is the distance between two particles, $i$ and $j$, $q_i$ and $q_j$ are the point charges of the two particles and $\varepsilon_0$ is the vacuum permittivity. The 12-6 LJ potential defines the energy with two parameters: $\varepsilon_{ij}$ (the well-depth of the potential) and $\sigma_{ij}$ (or $R_{\min,ij}$, the size parameter).

For a proper definition of the interaction potential the $\varepsilon_{ij}$ and $\sigma_{ij}$ parameters should be known for every possible $i$ and $j$ pairs. In the studied force field parameter sets either the ion-water (ion-oxygen) parameters ($\varepsilon_{iO}$ and $\sigma_{iO}$) or the ion $\varepsilon_{ii}$ and $\sigma_{ii}$ (or $R_{\min,ii}$) and the water $\varepsilon_{OO}$, $\sigma_{OO}$ are given and the unlike parameters can be calculated according to a given combination rule (in all the water models applied here $\varepsilon_{HH}=0$). The commonly used combination rules are the geometric combination rule and the Lorentz-Berthelot (LB) combination rule. For $\varepsilon_{ij}$ both of these use the geometric average:

$$\varepsilon_{ij} = \sqrt{\varepsilon_{ii}\varepsilon_{jj}} \ . \tag{3}$$

The $\sigma_{ij}$ is equal to the geometric average in the geometric combination rule:

$$\sigma_{ij} = \sqrt{\sigma_{ii}\sigma_{jj}} \tag{4}$$

According to the Lorentz-Berthelot combination rule $\sigma_{ij}$ is equal to the arithmetic average:

$$\sigma_{ij} = \frac{\sigma_{ii} + \sigma_{jj}}{2}. \tag{5}$$

$\varepsilon_{iO}$ and $\sigma_{iO}$ can be calculated applying the combination rules. One should be careful to use the proper combination rule (i.e. the one, specified in the original paper of the given FF), as was shown by Åqvist in an early paper [49].

Several of the investigated models focus on the ion-water interaction only, probably because most of them were calibrated by simulations in which a single ion was placed into water, thus only ion-water and water-water interactions were present. However in solution with finite (moderate or high) concentrations the ion-ion interactions are not negligible. The proper balance between the different ion-ion potentials (anion-anion, cation-cation and anion-cation) could be critical [13,41]. In most cases, however, the cation-anion parameters are not given separately. The users of these FFs apply one of the combination rules to calculate the cation-anion parameters. It has been recently demonstrated that this (mostly arbitrary) choice has, in some cases, a strong influence on the results of the simulation [50]. In this study I have applied the combination rule used for the calculation of the ion-water parameters to determine the cation-anion $\varepsilon$ and $\sigma$ values as well.

The ionic interatomic potentials will be discussed in detail in section 3. The $\varepsilon_{ii}$, $\sigma_{ii}$ and $q_i$ values for the models are collected in Table 1 together with the respective water model and the applied combination rule. The charges of the ions are $+1e$ for $Li^+$ and $-1e$ for $Cl^-$ in most of the force fields ($e$ is the electron charge). All of the calculated $\varepsilon_{ij}$ and $\sigma_{ij}$ parameters are given in the Supplementary Material (Table S.1). The investigated ionic interatomic potentials are developed for the simplest, and thus extensively used rigid, non-polarizable water models. These water models are: SPC/E [28], TIP4P [51] and TIP4PEw [52]. They are perhaps not the best rigid, non-polarizable water models (according to Ref. [53] the performance of the TIP4P/2005 [54] model is better than that of the SPC/E or TIP4P models); however, the ionic potential parameters were originally developed by adjusting to one of the above mentioned three water models. Since changing the water model alters the relevant $\varepsilon_{ij}$ and $\sigma_{ij}$ parameters, an appropriate investigation requires the application of the same water potential parameters that have been originally tested. Thus the performance of the ionic FFs with the original water models was analyzed here. (Some of the FF sets presented below were also determined in combination with the SPC [14] or TIP3P [51] water models (see section 3.), which water models perform even more poorly, thus they are not investigated here.) The SPC/E model is a 3-site model (the 3 coordinates are the position of the point-like oxygen and the two hydrogen atoms), while the others use a fourth, virtual site, which has a role in the charge distribution: the "charge of the oxygen atom" is located at the position of the virtual site. The parameters of the water models are collected in Table 2.

## 2.2 Simulation details

Classical molecular dynamics simulations were performed with the GROMACS software package (version 5.1.1) [55]. Simulation details were the same for all tested interatomic potentials. Usually the default settings (methods, algorithms, boundary conditions and parameters) of the GROMACS software were selected, as they are widely used in simulations.

Aqueous LiCl solutions were studied at four different concentrations; all of them were previously examined by neutron and X-ray scattering measurements [48]. Their molality values are between 3.74 and 19.55 mol/kg. The numbers of the atoms in the simulation boxes were around 10000. The exact number of water/ion pairs and the densities at 300 K are taken from Ref. [48] and shown in Table 3. The four studied LiCl solutions will be denoted throughout this work after their molality values as the 3.74m, 8.30m, 11.37m and 19.55m samples.

The simulations were carried out in cubic simulation boxes with periodic boundary conditions. The initial box sizes were calculated from the experimental densities and are shown in Table 3. The ions and water molecules were initially placed into the simulation box randomly. Water molecules were kept together rigidly by the SETTLE algorithm [56]. Van der Waals interactions were truncated at 10 Å, with added long-range corrections to energy and pressure [57]. Coulomb interactions were treated by the smoothed particle-mesh Ewald (SPME) method [58,59], using a 10 Å cutoff in the direct space, together with a grid spacing of 1.2 Å and fourth order polynomial for the interpolation of the reciprocal part. Different cutoff distances were also tested (15 Å, 20 Å). It was found that the investigated quantities do not depend on the cutoff length. The only exception is the static dielectric constant in those cases where the $\varepsilon(t)$ curves do not converge (see in sections 2.2.1 and 4.2).

At first the steepest-descent method (implemented in the GROMACS software) was used for energy minimization. After that the leap-frog algorithm was used for integrating Newton's equations of motion using a 2 fs time step. Equilibration was carried out at constant volume and temperature (NVT ensemble) at $T$= 300 K. During this first equilibration stage (200 ps) the Berendsen thermostat [60] was used with $\tau_T$=0.1 coupling, for relaxing the system to the target temperature value. After that the temperature was kept constant by the Nose-Hoover thermostat [61,62] with $\tau_T = 2.0$.

To reduce computational time and storage capacity requirements the same trajectories were used to calculate all investigated quantities, except the density. The simulations were conducted in the NVT ensemble, using box sizes determined from the experimental densities. Applying the same box sizes is favorable for structural investigations. The recommendations of Refs. [63, 64] were taken into account. The time step was 2 fs. Particle configurations were collected every 2 ps after a 2 ns equilibration period. The length of the (saved) trajectories was 8 ns.

*2.2.1 Static dielectric constant*

The static dielectric constant can be calculated according to Eq. (6):

$$\varepsilon = 1 + \frac{\langle M^2 \rangle - \langle M \rangle^2}{3\varepsilon_0 V k_B T} \quad ; \quad \text{(6)}$$

where $M$ is the total dipole moment, $\varepsilon_0$ is the vacuum permittivity, $V$ is the volume of the system, $k_B$ is Boltzmann's constant, and $T$ is the temperature. The calculations were carried out with the 'gmx dipole' program of the GROMACS software.

The average of the $\varepsilon$ values for the last 3 ns part of the $\varepsilon(t)$ curves (5ns-8ns) was determined and taken as final result.

*2.2.2 Self-diffusion coefficients*

The self-diffusion coefficients can be calculated from the mean-square displacement (MSD) or from the velocity auto-correlation function. According to Ref. [64] the first method is less sensitive to the length of the trajectories thus this method was used following the parameters of Ref. [64]. The self-diffusion coefficients, $D_A$ of particles type $A$ can be obtained from the mean square displacement by using the Einstein-relationship:

$$\lim_{t \to \infty} \left\langle \|r_i(t) - r_i(0)\|^2 \right\rangle_{i \in A} = 6D_A t . \quad \text{(7)}$$

The full trajectories were used by restarting the MSD-calculation every 10 ps. The linear part of the MSD curves, (the MSD-$t$ plot, without the first and last 10%) was fitted to determine $D_A$. The 'gmx msd' program of the GROMACS package was employed for the calculations.

*2.2.4 Structure factors*

For the determination of the structure factors every 80$^{th}$ saved configuration (in summary 50 configuration with a 160 ps time interval) was used. The 'gmx rdf' program of the GROMACS package was used to calculate the partial pair correlation functions (PPCF, $g_{ij}(r)$). The partial structure factors ($S_{ij}(Q)$) can be calculated from the PPCFs with equation (8):

$$S_{ij}(Q)-1 = \frac{4\pi\rho_0}{Q}\int_0^\infty r(g_{ij}(r)-1)\sin(Qr)dr, \qquad (8)$$

where $Q$ is the amplitude of the scattering vector, and $\rho_0$ is the average number density. The neutron and X-ray total structure factors can be expressed in terms of the partial structure factors as:

$$S(Q) = \sum_{i \leq j} w_{ij}^{X,N}(Q) S_{ij}(Q). \qquad (9)$$

Here $w_{ij}^{X,N}$ are the X-ray and neutron scattering weights. For neutrons they are given by equation (10):

$$w_{ij}^N = (2-\delta_{ij})\frac{c_i c_j b_i b_j}{\sum_{ij} c_i c_j b_i b_j}. \qquad (10)$$

Here $\delta_{ij}$ is the Kronecker delta, $c_i$ denotes the atomic concentration and $b_i$ is the coherent neutron scattering length. For the X-rays:

$$w_{ij}^X(Q) = (2-\delta_{ij})\frac{c_i c_j f_i(Q) f_j(Q)}{\sum_{ij} c_i c_j f_i(Q) f_j(Q)}, \qquad (11)$$

where $f_i(Q)$ is the atomic form factor. The weighting factors are shown in Fig. 1 for the most concentrated and the most dilute compositions. (It should be noted here that the similar figure, Fig. 2 of Ref. [25] is inaccurate concerning the labels of the XRD weights.)

*2.2.4 Density calculation*

During the density calculations the NpT ensemble was used, the pressure was kept at $p=10^5$ Pa by the Parrinello-Rahman barostat [65,66] using a coupling constant $\tau_p=2.0$. Densities were calculated every 20 ps from 2000 to 4000 ps, and the averages of the 100 values were determined.

## 3. Interatomic potential sets

-Chandrasekhar force field (Ch) [67]

This force field was calibrated to experimental interaction (binding) energies and calculated quantum-mechanical geometries (Hartree-Fock geometries) of ion-water complexes. TIP4P water and the geometric combination rule were used with this force field.

-Dang-Smith parameter set (DS) [68, 69]

Experimental gas-phase binding enthalpies for small ion-water clusters, measured structures of ionic solutions, and solvation enthalpies of ionic solutions were the targets during the parametrization of this model. This FF set was determined originally for (POL1 and RPOL) polarizable water (which have the same LJ parameters as SPC/E) [70, 71]. Later the parameters of the $Cl^-$ ion were modified for SPC/E water [69]. Parameters for $Li^+$ ions were also developed for POL1 water, later they were tested for SPC/E water as well [72-74]. The LB combination rule was used.

-Jensen-Jorgensen set (JJ) [75]

This FF was developed to obtain a consistent set for both anions and cations following the philosophy of the OPLS force field [76] and fitting the parameters to liquid-phase data: to experimental free energies of hydration and to the locations of the first maxima of the ion-oxygen radial distribution functions. In accordance with the practice for the OPLS FF, the geometric combination rule was used. The calculations were made with TIP4P water.

-Joung-Cheatham III (JC-S, JC-T) sets [4]

These parameters were fitted for hydration free energies, lattice energies and the lattice constants of salt crystals. Three different parameter sets are available in the publication, for TIP3P [51], TIP4PEw and SPC/E water models; the last two are examined in this study. The LB combination rule was used.

-Horinek-Mamatkulov-Netz parameter sets (HS-g, HM-g, HL-g, HS-LB, HM-LB, HL-LB) [22]

Solvation free energies and solvation entropies were the targets. It was found that for cations the problem is overdetermined and a set of parameters cannot be found which simultaneously fits both. The authors determined 3 σ parameters for 3 selected ε values (small (S), intermediate (M) and large (L) – 5, 5b, 5a in the original paper), which are fitted to the solvation free energies using a reference ion, $Cl^-$, which was parametrized by the DS parameters [69]. The water model was SPC/E. The ion-water $\sigma_{iO}$ and $\varepsilon_{iO}$ values were calculated, which leaves the choice of combination rule undetermined. As the cation-anion parameters were not calculated by the authors, I computed those with both of the two combination rules (which will be denoted in the name of these FF sets by '-g' for geometric and '-LB' for LB combination rule). Together with the three sets for the cations this results in 6 parameter sets.

-Gee et al. force field (Gee) [77]

This model was designed to reproduce the experimental Kirkwood-Buff integrals [78]; the lattice dimensions of sodium halide and alkali chloride crystals and ion-water contact distances were the target values. This parameter set was developed for SPC/E water using geometric (in some cases modified geometric) combination rules. For the $\varepsilon_{LiO}$ parameter a modified geometric combination rule was used: $\varepsilon_{LiO} = 0.4 \, (\varepsilon_{LiLi} \, \varepsilon_{OO})^{1/2}$.

-Reif, Hünenberger force fields [5] (RH, RM, RL, RL-sLB)

The homoionic $C_{6,II}$ parameters were obtained by the Slater-Kirkwood formula [79]; the $C_{12,IO}$ ion-oxygen parameters were fitted to experimental single-ion hydration free energies (See Eq. 2 for the correspondence of $C_6$ and $C_{12}$ with $\varepsilon$ and $\sigma$). The parameters were calculated for three different hydration free energy values (according to three absolute (intrinsic) hydration free energy of the proton; L-low, M-medium, H-high). The parameters were determined for SPC and SPC/E water models. The ion-oxygen parameters are given, the $C_{6,IO}$ ion-oxygen parameters were calculated from the homoionic $C_{6,II}$ and the $C_{6,OO}$ of the water-models by the geometric combination rule. I have tested these parameters with SPC/E water. The authors investigated some parameters of the ionic crystals as well and found that in the case of LiCl the L parameter set gives more reasonable results with applying the LB combination rule for the calculation of $\sigma_{LiCl}$. Thus the L set was investigated with a special LB rule as well (denoted as RL-sLB), in which $\sigma_{LiCl}$ is calculated according to the LB rule, but the $\sigma_{LiO}$ and $\sigma_{ClO}$ values remained the same as before.

-Mao-Pappu parameter set [80] (MP-S, MP-T)

A solvent-independent approach was used to calibrate the parameters, based exclusively on crystal lattice properties. The target values were the interionic distances and lattice energies of 20 cation-anion crystals (every possible combination of 5 cations and 4 anions). The force field parameters were optimized for these target values simultaneously. The LB mixing rule was used for the calculation of the parameters of unlike particles. As this model is solvent-independent it can be used with any water (and other solvent) models. Even though there may be better rigid, non-polarizable water models, I have tested this FF set only together with the SPC/E and TIP4PEw water models, since these are the water models used in the other cases. In this way the performance of the MP set can be compared to others.

-Deublein-Vrabec-Hasse force field [81] (DVH)

The $\sigma$ parameters were determined to obtain a good agreement with the reduced density (density of solution / pure water density) as a function of the solution salinity. The $\varepsilon$ parameters are fixed at a constant value for all ions. The authors calculated their parameters with SPC/E water, and claimed that it can be combined with other water models (such as TIP4P, or SPC etc.), as well. The LB combination rule was used.

-Reiser-Deublein-Vrabec-Hasse (RDVH) [82]

This parameter set is similar to the DVH set [81], with the same $\sigma$ values, the $\varepsilon$ values are determined on the basis of the experimental self-diffusion coefficients and the first peak of the ion-water radial distribution functions. The calculations were made with the SPC/E water model and the LB combination rule.

-Li-Song-Merz force fields (Li-HFE, Li-IOD) [6]

These authors defined two parameter sets. One of them (Li-HFE) focused on reproducing the experimental hydration free energies, the other one (Li-IOD) is fitted to the ion-oxygen distance of the first solvation shell (Li-IOD). They had found earlier that the 12-6 LJ model was unable to reproduce these two requirements simultaneously. (They proposed a 12-6-4 LJ model as well and parametrized it to reproduce the abovementioned values simultaneously; this model is not investigated here.) They optimized their parameter sets for three water models (TIP3P, SPC/E, and TIP4PEw); the last two sets are investigated here. The LB combination rule was used in these FF models.

*3.1 Modified force fields and their combinations used in recent publications*

-Åqvist + Chandrasekhar (AqCh) [67, 83]

The $Cl^-$ parameters of this combined FF are taken from the paper of Chandrasekhar [67] (see Ch FF above), which were parametrized with TIP4P water model and geometric combination rule. The $Li^+$ parameters are taken from the paper of Åqvist [83]. Åqvist fitted his force field parameters to obtain a good agreement for the hydration free energies. His $Li^+$ parameters are for the SPC water model and the geometric combination rule.

Although the $Li^+$ and $Cl^-$ ion parameters were originally developed for different water models, this combination is used by default when using the GROMACS software [55] in conjunction with the OPLS-AA force field. According to the OPLS-AA convention it is used with geometric combination rule.

This combination is used in several publications, such as Ref. [25] with different water models (SPC/E, TIP4P/2005 [54] and SWM4-DP [84]), Ref. [24] (with 5 different water models, none of them is TIP4P), or in Ref. [26] (with ethylene glycol).

This mixed ion set was tested in this study with SPC/E water.

-Pluharova et al. (Pl) [44, 45]

The authors of these two publications [44,45] investigated LiCl solutions at concentrations of 3 mol/kg and 6 mol/kg with MD simulations and neutron diffraction experiments. They tested some earlier LJ parameters (in a somewhat arbitrary way, e.g. the $\sigma$ and $\varepsilon$ values were taken from the OPLS-AA force field of GROMACS, but the LJ parameters of the unlike atoms were calculated with LB combination rule instead of the conventional geometric combination rule of the OPLS FF) and found poor agreement with experimental results. Thus they modified the sigma values of the $Li^+$ and $Cl^-$ ions. Moreover, to take into account the electronic polarizability of water in ionic solutions (electronic continuum correction, ECC), they reduced the ionic charges to $\pm 0.75e$.

The calculations were made with SPC/E water and LB combination rule, as seen in the supplementary information of Ref [44].

-Aragones et al. (Ar) [46]

In this publication the authors used the JC force field to investigate aqueous LiCl solutions in the 1.38 mol/kg – 9.25 mol/kg concentration range (in molality). They have found that this force field does not reproduce the experimental ion pairing correctly when the LB rules are used. They thus applied modified LB rules:

$$\varepsilon_{ij} = \chi \sqrt{\varepsilon_{ii}\varepsilon_{jj}} \tag{12}$$

$$\sigma_{ij} = \eta \frac{\sigma_{ii}+\sigma_{jj}}{2}. \tag{13}$$

Here $\chi = 1.88$ and $\eta = 0.932$ only for the Li-Cl pairs; for all other unlike atoms $\chi = 1$ and $\eta = 1$. The authors determined these values to obtain the best possible agreement with the "experimental" Li-Cl partial radial distribution functions of their least concentrated sample. The TIP4PEw water was used. (The authors checked the transferability of the JC FF parameters, originally calculated for TIP4PEw water, to the TIP4P/2005 water model, and observed that both FF produce the same results. The $\chi$ and $\eta$ values were determined for TIP4P/2005 water as well. This combination is not tested in this study.)

-Singh-Dalvi-Gaikar [27] (SDG-S, SDG-T)

These authors used a combination of the ionic parameters, where the so-called "OPLS-Li$^+$" parameters were similar to the DS parameters, and the "OPLS-Cl$^-$" parameters were equal with the JJ parameters. The LB combination rule was used with two water models: SPC/E and TIP4P water.

4. Results and discussion

To evaluate the quality of the investigated models the following properties were compared:
-solution density,
-static dielectric permittivity,
-self-diffusion coefficients,
-neutron and X-ray weighted structure factors,
which were examined for all four concentrations listed in Table 3.

*4.1 Density*

The densities obtained from the NpT simulations are presented in Fig. 2. (The values are also shown in Table S.2.) The experimental values from Ref. [48] are also shown for reference. The concentration dependence of the density is presented for some selected model in Fig. S.1.

For the lowest concentration investigated (3.74m) all models perform well: the obtained density values agree with the experimental density within 3%. As the concentration increases the agreement deteriorates: except for three FFs the values are lower than the experimental ones and the discrepancies have the tendency to increase (3-5% for 8.30m sample, 5-8% for 11.37m sample and 8-10% for the 19.55m sample, typically). The Ch, Gee, Li-IOD-S, Li-IOD-T, RL and RM models nevertheless give the best results for the density; the agreement with the experimental values remains within 5% for all concentrations. The performance of the RL model is particularly remarkable: the density values of this model agree with the experimental ones within 1%. It is worthwhile mentioning that the old Ch model also gives good agreement for the whole concentration region.

There are some surprising results as well. The DVH and RDVH models were parametrized to obtain good fit to the densities, but they are not among the best models in this investigation. Moreover, the densities obtained with these models are the lowest ones for the highest concentration. In the original paper (Fig. 4 in Ref. [81]) the concentration dependence of the solutions was investigated and a good agreement with the experimental values was found. Unfortunately LiCl solutions were examined only for low concentrations, the highest investigated concentration was around 3-4 mol/kg. In this concentration region (3.74m sample) the results presented here agree with the experimental value as well.

The values obtained with the RH model are in strong contrast with the other FFs: this model gives density values higher than the experimental ones for all concentrations. As it will be shown later, this model gives peculiar results also for the other investigated quantities (it has poor agreement with experimental values and/or behaves dissimilarly compared to the other FFs). The reason for this unusual behavior is clearly visible from the snapshots of the configurations. A representative configuration obtained with the RH model is shown in Fig. S.2 (for the 8.30m sample). A typical configuration of the 19.55m sample obtained with a more adequate FF is shown in Fig. S.3 for comparison. (The ion-ion PPCFs of these two configurations are also shown in Fig. S.4.) As it can be seen from the figures, LiCl precipitate is present in the configuration obtained with the RH model. A similar behavior was observed in all configurations obtained with the RH model: at the smallest investigated concentration as well as at higher concentrations.

*4.2 Static dielectric constant (static dielectric permittivity)*

The static dielectric constant was calculated in the same way for all models and all concentrations.

In Ref. [63] the authors investigated the effect of the trajectory length on the convergence of the static dielectric constant. They found that at least 6 ns long trajectories are necessary for the accurate determination of $\varepsilon$. Therefore, and to keep the total computational time at a reasonable value, 8 ns long trajectories were used here. As in the work referred to above, the last 3 ns part of the $\varepsilon(t)$ curves was used to determine the $\varepsilon$ values presented here.

The convergence of the $\varepsilon(t)$ curves was checked. It was found that for the 3.74 m sample all $\varepsilon(t)$ curves were converged within 8 ns. At higher concentrations some of the models produced $\varepsilon(t)$ curves which were slightly shifting even at 8 ns. At the highest concentration some curves are definitely not saturated. The convergence of these curves is highly dependent on the FF. Some $\varepsilon(t)$ curves are shown in Fig. 3.

The $\varepsilon$ values, determined as averages over the last 3 ns of the $\varepsilon(t)$ curves, are presented in Fig. 4 (and Table S.3). The concentration dependence of $\varepsilon$ for some models is shown in Fig. S.5. Values which are from definitely non-saturated curves are indicated by crosses. The concentration dependent experimental values, which were estimated according to Ref. [85], are also shown.

$\varepsilon$ values obtained from different FF models are scattered widely around the experimental values. For most models they are inside a ±25% region for the lowest investigated concentration. The discrepancy is higher for two models: RM and RH (32-36%). For the 8.30m sample the values are in a ±40% region, except for the above-mentioned two models, for which they are 65% higher than the experimental value. For the 19.55 m sample, several models give $\varepsilon$ values more than 50% smaller than the experimental one (DS, JC-S, JC-T, HM-g, HM-LB, Gee, DVH, and RDVH). The RH model gives a 50% higher value than the experimental one.

Overall, the Pl model gives the best agreement for the entire concentration range. The JJ and Ch models also give acceptable results. The Li-HFE-S, Li-HFE-T and Li-IOD-S results are promising but the convergence of the $\varepsilon(t)$ curves is questionable (at least at some concentrations).

## *4.3 Self-diffusion coefficients*

The self-diffusion coefficients of water and ions were calculated from the mean-square displacements, using the Einstein-relation (see Eq. (7)). The results are presented in Figs. 5-7 (and the values are also shown in Tables S.4, S.5 and S.6). The concentration dependence of the self-diffusion coefficients are shown in Fig. S.6 for some models. The experimental values were taken from Refs. [86, 87] or estimated by interpolation from the reported values. Nearly all simulated values are below the experimental ones. It was discussed in several papers that the self-diffusion coefficients calculated from simulations increase monotonously with increasing system size (see e.g. [64, 88]). The finite size correction according to Ref. [88] can be estimated by Eq. (12):

$$D = D_{\text{PBC}} + \frac{\xi k_B T}{6\pi\eta L}, \qquad (12)$$

where $D$ is the self-diffusion coefficient in the infinite size limit and $D_{\text{PBC}}$ is the self-diffusion coefficient calculated from a simulation with periodic boundary conditions. For a cubic simulation box, $\xi$=2.837297. $L$ is the box length and $\eta$ is the viscosity.

The size of this correction term, $D-D_{\text{PBC}}$, was calculated for the investigated samples. Experimental viscosities were used, which were taken from Ref. [89]. The calculated corrections are 9.2, 4.8, 3.0 and 0.85 × 10$^{-11}$ m$^2$/s for the 3.74m, 8.30m, 11.37 and 19.55m samples, respectively. These corrections are significantly smaller than the discrepancies between the simulated and the experimental results. The experimental values reduced by the correction terms are also shown in Figs. 5-7.

The simulated chloride ion self-diffusion coefficients are smaller than the experimental values for all models except the Pl model in the entire investigated concentration range. The difference is more than 30% even for the 3.74m sample and it is increasing as concentration increases.

A similar trend can be observed for the concentration dependency of the simulated $D_{Li}$ values. The predictions of the models for $D_{Li}$ are better than for $D_{Cl}$ in the case of the 3.74m sample. For higher concentrations the results are equally inadequate. The Pl model gives significantly higher values than the other models, even higher than the experimental ones.

It should be noted here that the ionic self-diffusion coefficients (both $D_{Li}$ and $D_{Cl}$) obtained for the 19.55m sample are extremely low, except for the Pl model. This means that for all these models the Li$^+$ and Cl$^-$ ions essentially do not move, which makes these models unphysical from the point of view of ionic self-diffusion.

The simulated water self-diffusion coefficients are equal to the corrected experimental values within 10% for several models in the case of the 3.74m sample. As concentration increases, results become more scattered. For the 8.3m sample 6 models give values within the ±15% region (Li-IOD-T, Li-IOD-S, RM, AqCh, RL, SDG-S), for the 11.37m sample only four do (Li-IOD-T, Li-IOD-S, RM, Ch). In the case of the highest concentration there are only 3 models which give results equal to the corrected experimental value within ±30%: JJ, Li-IOD-T and RM. The Li-IOD-T and RM models can be said to provide the best results considering the entire investigated concentration range.

Two models give higher values than the experimental ones for all samples: the RH and the Pl models. As was discussed previously, there is LiCl precipitate in the configurations obtained with the RH model.

The behavior of the Pl model is unique among the investigated models: it gives higher values than the experimental ones for the ionic and for the water self-diffusion coefficients. The origin of this dissimilarity can be that in this model the ion charges are rescaled (they are ±0.75$e$). The electronic continuum correction method [90], applied in the Pl model, has also been used by others with promising results (see e.g. Refs. [43, 91-95] and references therein). A water model dependent but ionic model independent charge rescaling method was proposed recently in Ref. [94]. The authors found that their results for self-diffusion coefficients with the rescaled charges are significantly better than those of the original models. (I have also checked some models (JC-T, MP-T, RL) with rescaled charges and found a similar tendency. Moreover, the static dielectric constant values of the models with rescaled charges were also better than the original ones for this model. Unfortunately the differences for the density values were higher. It can be the subject of a follow-up study to expand these investigations to all of the models studied here.)

*4.4 Neutron and X-ray weighted structure factors*

Neutron and X-ray diffraction measurements are the most widely used experimental techniques to obtain structural information. The total scattering structure factor (for neutrons ($S^N(Q)$), for X-rays ($S^X(Q)$)) can be calculated from the measured intensities after the necessary corrections. The relationship between the structure factors and the partial pair correlation functions are described in Eqs. (8-11).

Experimental neutron and X-ray structure factors of LiCl solutions were taken from Ref. [48]. (Their neutron diffraction experiments were performed using deuterium instead of hydrogen.) The calculated $S^N(Q)$ and $S^X(Q)$ functions for selected FFs are shown in Figs. 8 and 9. To compare the quality of these fits with experiment and with each other, the *R*-factors ('goodness-of-fit' values) were calculated according to Eq. (13):

$$R = \frac{\sqrt{\sum_i (S_{\mathrm{mod}}(Q_i) - S_{\mathrm{exp}}(Q_i))^2}}{\sqrt{\sum_i (S_{\mathrm{exp}}(Q_i) - 1)^2}}. \qquad (13)$$

Here $S_{\mathrm{mod}}$ and $S_{\mathrm{exp}}$ are the model and experimental structure factors, and the summation is over the $i$ experimental points. The *R*-factors of the neutron and X-ray structure factors calculated for the different FFs are presented in Figs. 10 and 11 (and Tables S.7 and S.8). (I note that for the calculations of the $S^N(Q)$ and $S^X(Q)$ structure factors the total (intra and intermolecular) O-H and H-H PPCFs are used. The molecular geometry of the applied water model can influence the quality of the fits, see e. g. Ref. [96].)

The neutron weighted total structure factor is reproduced reasonably well by most models except for the highest concentration. In the case of the lowest concentration the best fits are obtained in conjunction with the TIP4PEw water potential (JC-T, Li-HFE-T, Li-IOD-T, MP-T, Ar models). Results with the TIP4P water model (Ch, JJ, SDG-T) are also superior to the ones with the SPC/E water model. This can be well understood from the values of the neutron scattering weights (see Fig. 1). The neutron diffraction experiment is sensitive mostly to the H-H (D-D) and O-H (O-D) pair correlations for all concentrations. At the highest concentration the contribution of the Cl-H (Cl-D) partial is also important.

As concentration increases, differences between the FFs become more significant. For the 19.55m sample only three of the five models with TIP4PEw water model lead to adequate fits with the experimental neutron structure factor, JC-T, MP-T and Ar. The JJ, Li-IOD-T and Li-HFE-T models become worse as concentration increases. Among the models developed for the SPC/E water model, Pl gives the best fit with the experimental neutron structure factor.

The X-ray structure factors are reproduced less successfully by the FFs investigated. In this function, contributions of ion-ion and ion-water type partials are more significant than in the neutron total structure function. The weights of the O-O and Cl-O partials are the highest for the most dilute solution and the contribution of the O-H partial is also important for the low $Q$ region. For the highest concentration, beside the Cl-O and O-O partials, the Cl-Cl partial plays an essential role in the total structure factor, as well. The Li-Cl, Li-O and Cl-H partials (together with the O-H) are appreciable in the low $Q$ region.

In this test the JC-S and JC-T models were the best, but the quality of the fits is far from satisfactory for the 19.55m sample. The MP-S, MP-T, Ar, Pl and Gee FF models are somewhat better than the remaining ones. The results for the X-ray structure factor of the JJ, Li-IOD-T, and SDG-T models are the worst. The AqCh model, which was used in Ref. [25], also gives poor agreement with the experimental $S^X(Q)$ in the entire investigated concentration range.

The Ar model, which is based on the JC-T model and has the Li-Cl LJ cross-parameters from a modified LB rule, was developed to obtain a better agreement with the Li-Cl partial pair correlation function. The result of this effort can be seen in Fig. 9: the quality of the fit obtained with the Ar model is indeed better than that of the JC-T model for the lowest concentration. Unfortunately, for the highest concentration this refinement is not successful and $S^X(Q)$ obtained with the original JC-T model is slightly better than that of the Ar model.

Taking into account the poor agreement of the simulated curves with the experimental ones, the structural predictions of the MD models can be questioned. The consistency of the calculated partial pair correlation functions with the experimental structure factors will be further analyzed in a follow-up publication, together with a detailed analysis of the structural information obtained from different FF models. Here I would like to discuss only the first maxima of the ion-oxygen partial pair correlation functions. This quantity is often used as target for the determination of FF parameters or as a control parameter during while testing the models.

The Li-O PPCF shows the hydration shell of the Li$^+$ ions. I found that for all FFs the position of the first maxima of this curve does not change with concentration. The average values are around 1.96-2.02 Å for most FFs, see Fig. 12 (and Table S.9). There are a few exceptions: the Li-O nearest neighbor distance is significantly shorter with the Gee (1.88 Å) and Li-HFE-T (1.9 Å) models, and longer for the RH (2.18 Å), RM (2.08 Å), DVH (2.2 Å), RDVH (2.25 Å) and Li-IOD-T (2.08 Å) models. The results presented here are in good agreement with the values reported in the original papers.

The Cl-O distances are less well defined and for several models they strongly depend on the concentration. These models are: HS-g, HM-g, HL-g, HS-LB, RL, RDVH, Li-HFE-S, AqCh, Ar and SDG-T. A sensitivity of the Cl-O distances to counterions and concentration has been observed earlier [97]. In these models shorter values correlate with the lower concentrations, the Cl-O distances at the highest concentration are 0.1-0.25 Å longer than at the lowest one. The simulations for the other FFs give values, which depend only slightly on concentration (or not at all). The Cl-O distances scatter in the 2.96 Å (RH) to 3.50 Å (JJ) range, see Fig. 12 (and Table S.10). Most of them are similar to the results obtained by the developers of the FFs. Higher values are found here for the Ch (3.34 Å instead of 3.21 Å) and JJ (3.50 Å instead of 3.25 Å) models. For the models where the Cl-O distance depends on the concentration, the lowest calculated values are similar to the originally reported ones.

*4.5 Further remarks*

According to Ref. [82] the RDVH model is a refinement of the DVH model, in which the epsilon parameters are determined to obtain a better agreement with experimental results for the ion-oxygen distances and the self-diffusion coefficients. The results obtained here did not confirm this: the self-diffusion values obtained with the RDVH FF are somewhat smaller (and worse) than those for the DVH model. The $R$-factors of the simulated $S^N(Q)$ and $S^X(Q)$ curves are higher for the RDVH than for the DVH model, suggesting poorer structural agreement. The Li-O distance is higher for the RDVH model (it is the highest value) and the Cl-O distance obtained with RDVH depends on the concentration.

As the FFs are parametrized altogether with 12 LJ parameters (O-O, Li-O, Cl-O, Li-Li, Cl-Cl and Li-Cl sigma and epsilon values) it is hard to find direct relationship between the investigated quantities and the LJ parameters. However some correspondence between them can be observed.

The role of the ion-ion (cation-anion) LJ parameters can be studied by comparing the results of the different Horinek-Mamatkulov-Netz parameter sets [22]. The authors determined the ion-oxygen parameters only; the ion-ion parameters can be calculated according to the combination rule, which rule was left undisclosed. Applying different combination rules results in different ion-ion LJ $\sigma$ parameters (the calculation of the $\varepsilon$ parameters are the same for the LB and the geometric combination rules). The effect of the selected combination rule (and thus the values of the ion-ion parameters) can be investigated by comparing their FF sets with different combination rules (the HS-g model can be compared to the HS-LB, HM-g to HM-LB, and HL-g to HL-LB). The $\sigma_{LiLi}$ and $\sigma_{ClCl}$ parameters are higher, while the $\sigma_{LiCl}$ parameters are lower for the models with geometric combination rule than the corresponding values for the models with LB rules (see Table S.1 in the Supplementary Material). The differences are small for the HS sets (0.3%, 2.7% and 0.8% for $\sigma_{LiLi}$, $\sigma_{ClCl}$ and $\sigma_{LiCl}$, respectively), while they are higher for the HM (16%, 2.7% and 6%) or HL (19%, 2.7%, 6.6%) sets.

The results of the HS-g and HS-LB models are similar, their differences are not significant. The densities and water self-diffusion coefficients are smaller (and thus "worse") for HM-LB and HL-LB than for HM-g and HL-g. The opposite is true for the ionic self-diffusion coefficients. The static dielectric constants obtained with models using the LB rules are always smaller than those obtained with models with the geometric combination rules, but some of the $\varepsilon(t)$ curves (for the higher concentrations) did not converge. The $R$-factors for HM-LB and HL-LB are slightly better for $S^N(Q)$ and markedly better for $S^X(Q)$ curves. The Cl-O distances are smaller for the HM-LB and HL-LB models than for the HM-g and HL-g ones.

A similar comparison between RL and RL-sLB (where only the $\sigma_{LiCl}$ is different: slightly higher for RL-sLB) shows that the ionic self-diffusion values are higher (and closer to the experimental values) and the X-ray diffraction fits are better for RL-sLB. In the RL model, where the $\sigma_{LiCl}$ value is smaller, the water-self diffusion constant and the static dielectric constant are higher. The densities and neutron diffraction fits are similar.

The Ar and JC-T models differ only in the $\sigma_{LiCl}$ and $\varepsilon_{LiCl}$ parameters (the former is 7% higher in the JC-T model, the latter one is 88% higher in the Ar model). The simulations with these two models show that the alterations made in the Ar models improve the results of the self-diffusion coefficients (mostly the water self-diffusion coefficient), the static dielectric constant and somewhat the density. The static dielectric constant and the water self-diffusion constant are higher for the Ar model (which has lower $\sigma_{LiCl}$), than for the JC-T FF. This result is consistent with that found for the RL and RL-sLB models and also for the HM and HL models.

The RH model (which model led to configurations in which precipitation was observed) has the highest $\sigma_{LiLi}$ and $\sigma_{LiO}$ parameters and the lowest $\sigma_{ClCl}$ and $\sigma_{ClO}$ values among the FFs. The $\sigma_{LiO}/\sigma_{ClO}$ ratio (or the $\sigma_{LiLi}/\sigma_{ClCl}$ ratio) is around 1 in the RH model, while this value is between 0.5 and 0.8 for most FFs. The smallest ionic self-diffusion constants are obtained with the JJ, RH and RM models, where the $\sigma_{LiO}/\sigma_{ClO}$ (or the $\sigma_{LiLi}/\sigma_{ClCl}$) ratio is the highest. This correspondence between the $\sigma_{LiLi}/\sigma_{ClCl}$ ratio and the ionic self-diffusion constants is in agreement with the observations for the HM and HL FFs. The $\sigma_{LiLi}/\sigma_{ClCl}$ ratios are 0.376 and 0.334 for HM-g and HM-LB, and 0.361 and 0.311 for HL-g and HL-LB. The ionic self-diffusion coefficients are higher for HM-LB and HL-LB, than for HM-g and HL-g.

The quality of the $S^X(Q)$ fit seems to be sensitive to the $\varepsilon_{ClO}$ values. The models with lower $\varepsilon_{ClO}$ values give better agreement with the experimental structure factor than those models where this value is higher (for JC-T, JC-S and Ar models $\varepsilon_{ClO} \approx 0.18$, for MP-T and MP-S $\varepsilon_{ClO} \approx 0.26$, for JJ, Li-IOD-T, SDG-T $\varepsilon_{ClO}$ is higher than 1.2).

The role of the water model can be tested by comparing results obtained with the MP-S and MP-T models. This clearly shows that the TIP4PEw water model is more suitable than the SPC/E model: all except one investigated quantities show a better agreement with experimental results for the MP-T model than for the MP-S model. (The only exception is the *R*-factor of the X-ray structure factor for the highest concentration.) One can find several other water models in the literature. Testing all, or at least more of them with the ionic FF presented here, is beyond the scope of this work. The comparison of the MP-S and MP-T models suggest that an appropriate water model can improve the performance of the FF. Similar results were obtained recently for aqueous NaCl solutions [98], where the JJ and JC-T FF-s were checked with a new water model, namely TIP4P/ε [99]. This water model was developed using static dielectric constant at room temperature and the temperature of maximum density as target properties. It was found that the new combinations give better agreement with the experimental values about the static dielectric constant and the self-diffusion coefficients for NaCl solutions.

## 5. Summary and conclusions

Aqueous lithium chloride solutions up to high concentrations were studied by classical molecular dynamics simulations. 29 interaction models were examined; all use 12-6 LJ and Coulomb interactions between the particles with rigid, non-polarizable water molecules. Their predictions about the density, the static dielectric constant and (ionic and water) self-diffusion coefficients were obtained for four concentrations from 3.74 mol/kg to 19.55 mol/kg. The neutron and X-ray weighted structure factors were determined and results for the various force fields were compared to experimental data.

A summary of the performances of the models, considering all investigated properties, is presented in Table 4. Models which are significantly better or worse than others according to one of the tested quantities are marked.

It was found that the usefulness of these simple models for highly concentrated aqueous LiCl solutions is limited. None of the investigated models gives satisfactory results for all tested quantities. The best performing model is the one presented by Pluharova *et al.* [44, 45], which model was developed directly for moderately and highly concentrated LiCl solutions and uses reduced charges. Two sets by Reif and Hünenberger [5] and two sets by Li, Song and Merz [6], namely RM, RL, Li-IOD-S and Li-IOD-T are promising concerning the density and water self-diffusion coefficients. The models by Joung and Cheatham III [4] (JC-S and JC-T) and the parameters by Mao and Pappu [80] (MP-S and MP-T) are the most suitable ones to gain structural information.

The simulations verified that refitting the cation-anion parameters by using modified combination rules as was done by Aragones *et al.* [46], – which is equivalent to determining not only the ion-water but the anion-cation parameters separately – is a useful and necessary method to enhance the performance of 12-6 LJ FFs. In most of the FFs only 6 $\sigma_{ij}$, $\varepsilon_{ij}$ parameters are determined independently ($\sigma_{OO}$, $\varepsilon_{OO}$, $\sigma_{LiO}$, $\varepsilon_{LiO}$, $\sigma_{ClO}$, $\varepsilon_{ClO}$) and the remaining $\sigma_{ij}$, $\varepsilon_{ij}$ values are declared with an arbitrary chosen combination rule. Using independent $\sigma_{ij}$, $\varepsilon_{ij}$ parameters for each type of atom pairs (*i-j*) (refitting $\sigma_{LiLi}$, $\varepsilon_{LiLi}$, $\sigma_{ClCl}$, $\varepsilon_{ClCl}$, $\sigma_{LiCl}$, $\varepsilon_{LiCl}$ also) can significantly improve the usability of the FFs. Such a refinement is possible, if instead of a few target parameters, a significant portion of the available data is taken into account during the creation and initial evaluation of the models. The current study is intended to provide a possible starting point for such efforts.

Correlations between some parameters and investigated quantities were observed. The $\sigma_{LiLi}/\sigma_{ClCl}$ ratio seems to be inversely proportional to the ionic self-diffusion constants, while lower $\sigma_{LiCl}$ leads to higher water self-diffusion constant and higher static dielectric constant. The $\varepsilon_{ClO}$ value is an important parameter concerning the X-ray diffraction structure factor: decreasing the $\varepsilon_{ClO}$ parameter decreases the *R*-factor of the $S^X(Q)$ fit. The effect of the water model, which can be well observed in the fit of the neutron diffraction structure function, is diminishing as the concentration of the solution increases.

The calculations confirmed that the charge rescaling method, used by Pluharova *et al.* improves the agreement with experimental data, primarily for the static dielectric constant and the self-diffusion coefficients.

To obtain accurate results about highly concentrated aqueous LiCl solutions from classical molecular dynamics simulations it is necessary to construct more appropriate 12-6 LJ FFs by combining the above mentioned methods (charge rescaling, independent LJ parameters) or to apply more complex models (such as the 12-6-4 LJ model [18] or polarizable ion FFs (e.g. AH/SWM4-NDP [20], AH/BK3 [21]).

Acknowledgments

The author is grateful to the National Research, Development and Innovation Office (NKFIH) of Hungary for financial support through Grant No. SNN 116198. The author would like to thank L.

**Table 1**

Force field parameters of the investigated models. The applied water models and combination rules are also given. For the definitions of the combination rules see text (mgeom means modified geometric rule, mLB means modified LB rule).

| Model | $\sigma_{LiLi}$ [nm] | $\varepsilon_{LiLi}$ [kJ/mol] | $\sigma_{ClCl}$ [nm] | $\varepsilon_{ClCl}$ [kJ/mol] | $q_{Li}/q_{Cl}$ [e] | comb. rule | water model | Ref. |
|---|---|---|---|---|---|---|---|---|
| Ch | 0.126 | 26.1495 | 0.4417 | 0.4928 | +1/-1 | geom | TIP4P | [67] |
| DS | 0.1506 | 0.6904 | 0.4400 | 0.4184 | +1/-1 | LB | SPC/E | [68,69] |
| JJ | 0.2870 | 0.0021 | 0.4020 | 2.9706 | +1/-1 | geom | TIP4P | [75] |
| JC-S | 0.1409 | 1.4089 | 0.4830 | 0.0535 | +1/-1 | LB | SPC/E | [4] |
| JC-T | 0.1440 | 0.4351 | 0.4918 | 0.0488 | +1/-1 | LB | TIP4PEw | [4] |
| HS-g | 0.2880 | 0.0006 | 0.4520 | 0.4200 | +1/-1 | geom | SPC/E | [22] |
| HM-g | 0.1700 | 0.6500 | 0.4520 | 0.4200 | +1/-1 | geom | SPC/E | [22] |
| HL-g | 0.1630 | 1.5400 | 0.4520 | 0.4200 | +1/-1 | geom | SPC/E | [22] |
| HS-LB | 0.2870 | 0.0006 | 0.4400 | 0.4200 | +1/-1 | LB | SPC/E | [22] |
| HM-LB | 0.1470 | 0.6500 | 0.4400 | 0.4200 | +1/-1 | LB | SPC/E | [22] |
| HL-LB | 0.1370 | 1.5400 | 0.4400 | 0.4200 | +1/-1 | LB | SPC/E | [22] |
| Gee | 0.182 | 0.7 | 0.44 | 0.47 | +1/-1 | mgeom | SPC/E | [77] |
| RH | 0.3529 | 0.0007 | 0.3493 | 1.7625 | +1/-1 | geom | SPC/E | [5] |
| RM | 0.3078 | 0.0015 | 0.3771 | 1.1137 | +1/-1 | geom | SPC/E | [5] |
| RL | 0.2679 | 0.0035 | 0.4096 | 0.6785 | +1/-1 | geom | SPC/E | [5] |
| RL-sLB | 0.2679 | 0.0035 | 0.4096 | 0.6785 | +1/-1 | sLB | SPC/E | [5] |
| MP-S | 0.1715 | 0.2412 | 0.4612 | 0.1047 | +1/-1 | LB | SPC/E | [80] |
| MP-T | 0.1715 | 0.2412 | 0.4612 | 0.1047 | +1/-1 | LB | TIP4PEw | [80] |
| DVH | 0.1880 | 0.8314 | 0.4410 | 0.8314 | +1/-1 | LB | SPC/E | [81] |
| RDVH | 0.1880 | 1.6629 | 0.4410 | 1.6629 | +1/-1 | LB | SPC/E | [82] |
| Li-HFE-S | 0.2242 | 0.0115 | 0.4112 | 2.6931 | +1/-1 | LB | SPC/E | [6] |
| Li-HFE-T | 0.2184 | 0.0071 | 0.4136 | 2.7309 | +1/-1 | LB | TIP4PEw | [6] |
| Li-IOD-S | 0.2343 | 0.0249 | 0.3852 | 2.2240 | +1/-1 | LB | SPC/E | [6] |
| Li-IOD-T | 0.2343 | 0.0249 | 0.3852 | 2.2240 | +1/-1 | LB | TIP4PEw | [6] |
| AqCh | 0.2126 | 0.0765 | 0.4417 | 0.4928 | +1/-1 | geom | SPC/E | [67,83] |
| Pl | 0.1800 | 0.0765 | 0.4100 | 0.4928 | +0.75/-0.75 | LB | SPC/E | [44,45] |
| Ar | 0.1440 | 0.4351 | 0.4918 | 0.0488 | +1/-1 | mLB | TIP4PEw | [46] |
| SDG-S | 0.1506 | 0.6945 | 0.402 | 2.9706 | +1/-1 | LB | SPC/E | [27] |
| SDG-T | 0.1506 | 0.6945 | 0.402 | 2.9706 | +1/-1 | LB | TIP4P | [27] |

**Table 2**

Parameters of the water-models. In the TIP4P and TIP4PEw models there is a fourth (virtual) site (M). It is situated along the bisector of the H-O-H angle and coplanar with the oxygen and hydrogens. The negative charge is placed in M.

| | $\sigma_{OO}$ [nm] | $\varepsilon_{OO}$ [kJ/mol] | $q_H$ [e] | $d_{O-H}$ [nm] | $\theta_{H-O-H}$ [deg] | $d_{O-M}$ [nm] | Ref. |
|---|---|---|---|---|---|---|---|
| SPC/E | 0.3166 | 0.6502 | +0.4238 | 0.1 | 109.47 | - | [28] |
| TIP4P | 0.3154 | 0.6485 | +0.52 | 0.09572 | 104.52 | 0.015 | [51] |
| TIP4PEw | 0.3164 | 0.6809 | +0.52422 | 0.09572 | 104.52 | 0.0125 | [52] |

**Table 3**

Investigated concentrations. The numbers of ion pairs/water and the experimental densities are taken from Ref. [48].

| $m$ [mol/kg] | 3.74 | 8.30 | 11.37 | 19.55 |
|---|---|---|---|---|
| $N_{LiCl}$ | 200 | 500 | 700 | 1000 |
| $N_{water}$ | 2968 | 3345 | 3416 | 2840 |
| Density [g/cm$^3$] | 1.076 | 1.1510 | 1.1950 | 1.2862 |
| Number density [Å$^{-3}$] | 0.09735 | 0.0939 | 0.0919 | 0.0871 |
| Box length [nm] | 4.5721 | 4.8982 | 5.0232 | 4.94102 |

**Table 4**

Comparison of the investigated FF models, considering 7 properties: density, static dielectric constant, self-diffusion coefficients of Li$^+$ ions, Cl$^-$ ions and water molecules, neutron and X-ray weighted total structure factors. The models, which give better result about the investigated quantity than the average, are marked with sign +, the best of them with sign ++. The models, which give worse result about the investigated quantity than the average are marked with sign -, the worst of them with sign --. Question mark denotes models in which not properly saturated $\varepsilon$ values cause uncertainties.

|          | $\rho$ | $\varepsilon$ | $D_{Li}$ | $D_{Cl}$ | $D_{water}$ | $S^N(Q)$ | $S^X(Q)$ |
|----------|--------|---------------|----------|----------|-------------|----------|----------|
| Ch       | +      | +             |          | -        | +           | +        |          |
| DS       | -      | -             |          | -        |             |          |          |
| JJ       |        | +             | -        | -        |             | --       | --       |
| JC-S     |        | -             |          | -        |             |          | ++       |
| JC-T     |        | -             |          | -        |             | ++       | ++       |
| HS-g     |        | --?           | -        |          | -           |          |          |
| HM-g     |        | -             | -        |          | -           |          |          |
| HL-g     |        | --?           | -        |          | -           |          | -        |
| HS-LB    |        |               | -        |          | -           |          |          |
| HM-LB    | -      |               | -        | -        | -           |          |          |
| HL-LB    |        |               |          |          | -           |          |          |
| Gee      | +      | -             | -        | -        |             |          | +        |
| RH       | --     | --            | -        | -        |             | --       | -        |
| RM       | +      | -             | -        | -        | ++          |          | -        |
| RL       | ++     | -             | -        | -        | +           |          |          |
| RL-sLB   |        |               |          |          |             |          |          |
| MP-S     |        |               |          | -        |             |          | +        |
| MP-T     |        |               |          | -        |             | ++       | +        |
| DVH      | --     | -             |          | -        |             | -        | -        |
| RDVH     | --     | -             |          | -        |             | --       | --       |
| Li-HFE-S | -      | +?            |          | -        |             | -        | -        |
| Li-HFE-T |        | +?            |          | -        |             |          | -        |
| Li-IOD-S | +      | +?            | -        | -        | ++          | -        | -        |
| Li-IOD-T | +      |               | -        | -        | ++          |          | --       |
| AqCh     |        | ?             | -        | -        | +           |          | -        |
| Pl       | -      | ++            | +        | ++       | +           | +        | +        |
| Ar       |        |               | -        | -        |             | ++       | +        |
| SDG-S    |        |               |          | -        | +           |          | -        |
| SDG-T    |        |               |          | -        |             |          | --       |

Figures

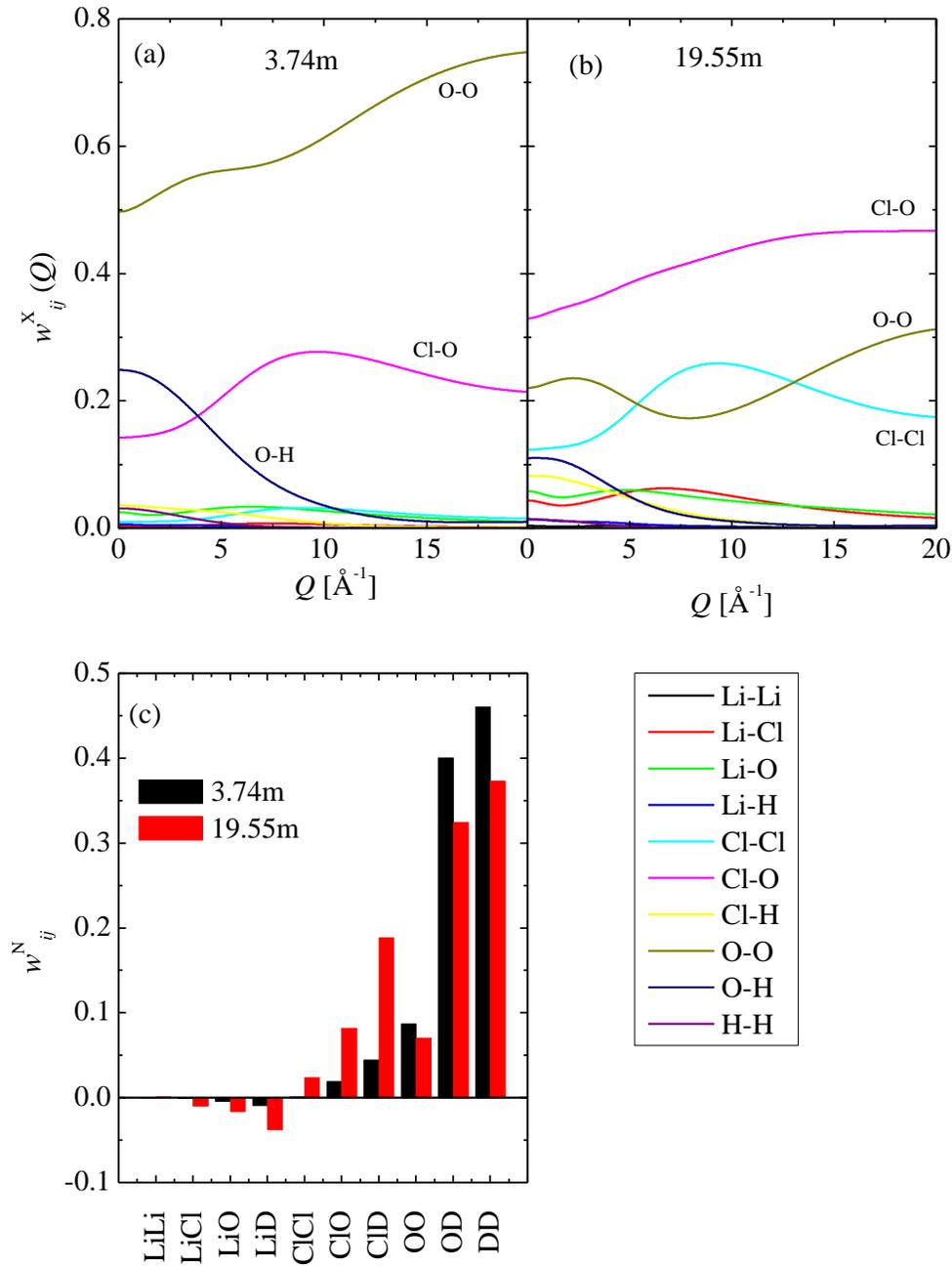

**Fig. 1** X-ray (a,b) and neutron (c) scattering weights used for the calculations of the X-ray and neutron total structure factors. Both quantities are shown for the most dilute (3.74m) and for the most concentrated (19.55m) samples. (The X-ray weighting factors are $Q$-dependent.)

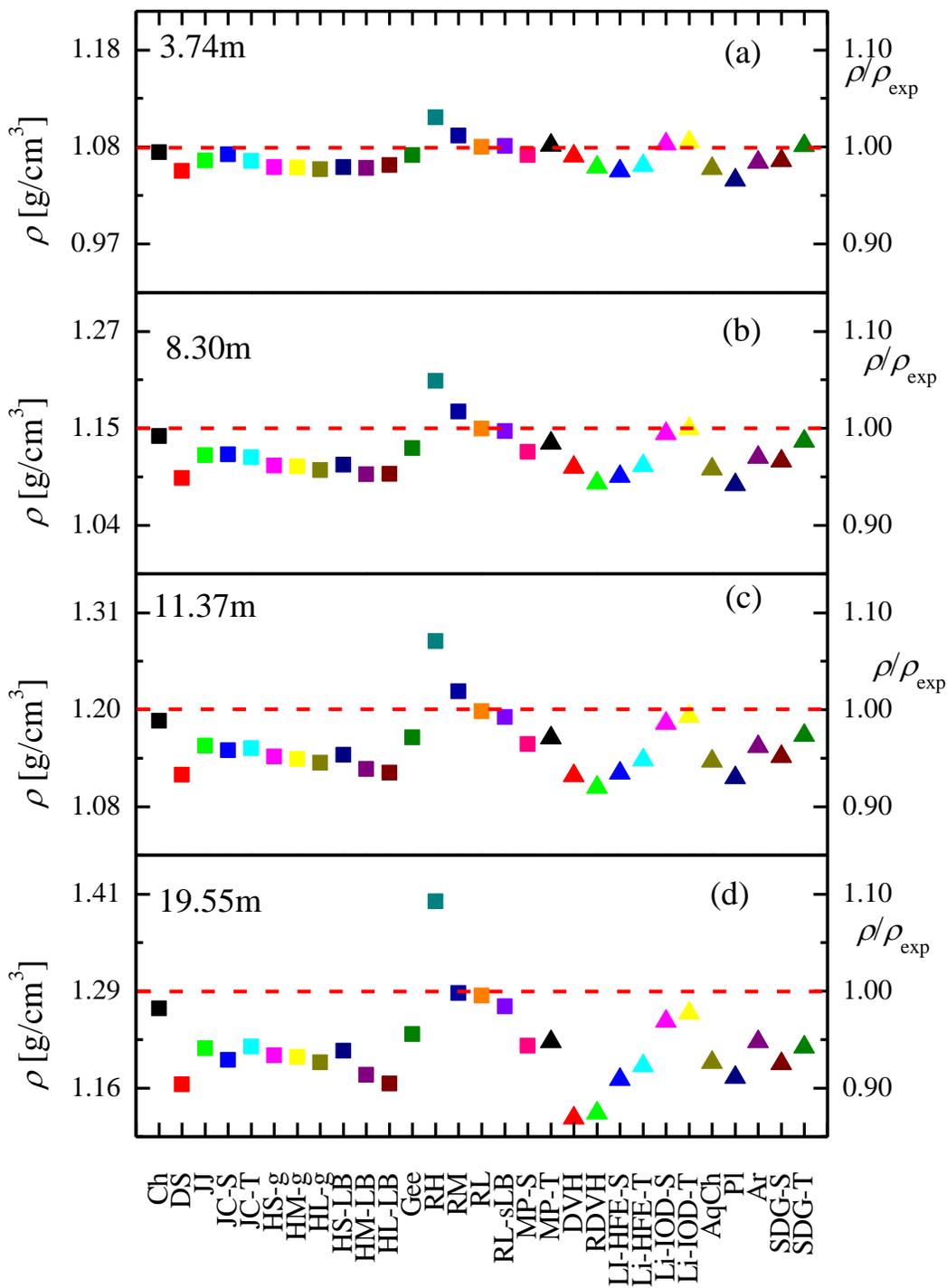

**Fig. 2** Densities at the concentrations (a) 3.74 mol/kg, (b) 8.30 mol/kg, (c) 11.37 mol/kg, and (d) 19.55 mol/kg, obtained from the simulations with different FFs. Experimental values are marked with dashed lines. The ratios of the simulated and experimental values ($\rho/\rho_{exp}$) are given by the ordinate on the right hand side.

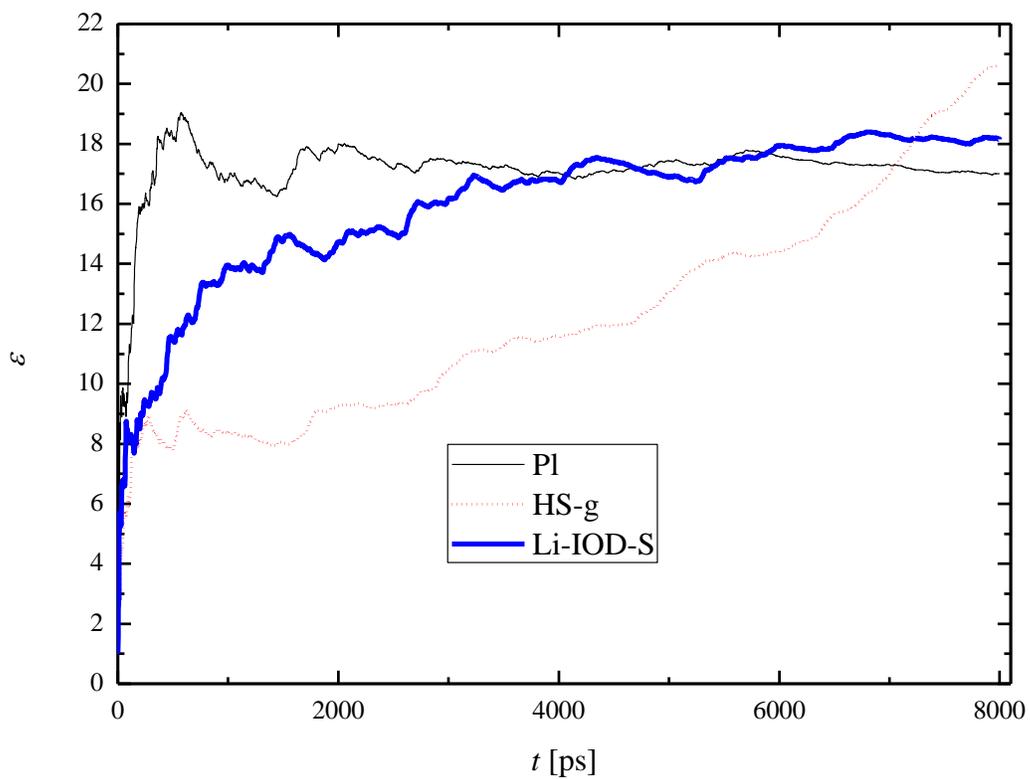

**Fig. 3** Convergence of the static dielectric constant ($\varepsilon$) for three selected models (Pl, HS-g and Li-IOD-S) at the concentration $m$=19.55 mol/kg. The curve is converged for the Pl model, still slightly evolving for the Li-IOD-S model, and definitely not converged even at 8 ns, for the HS-g model.

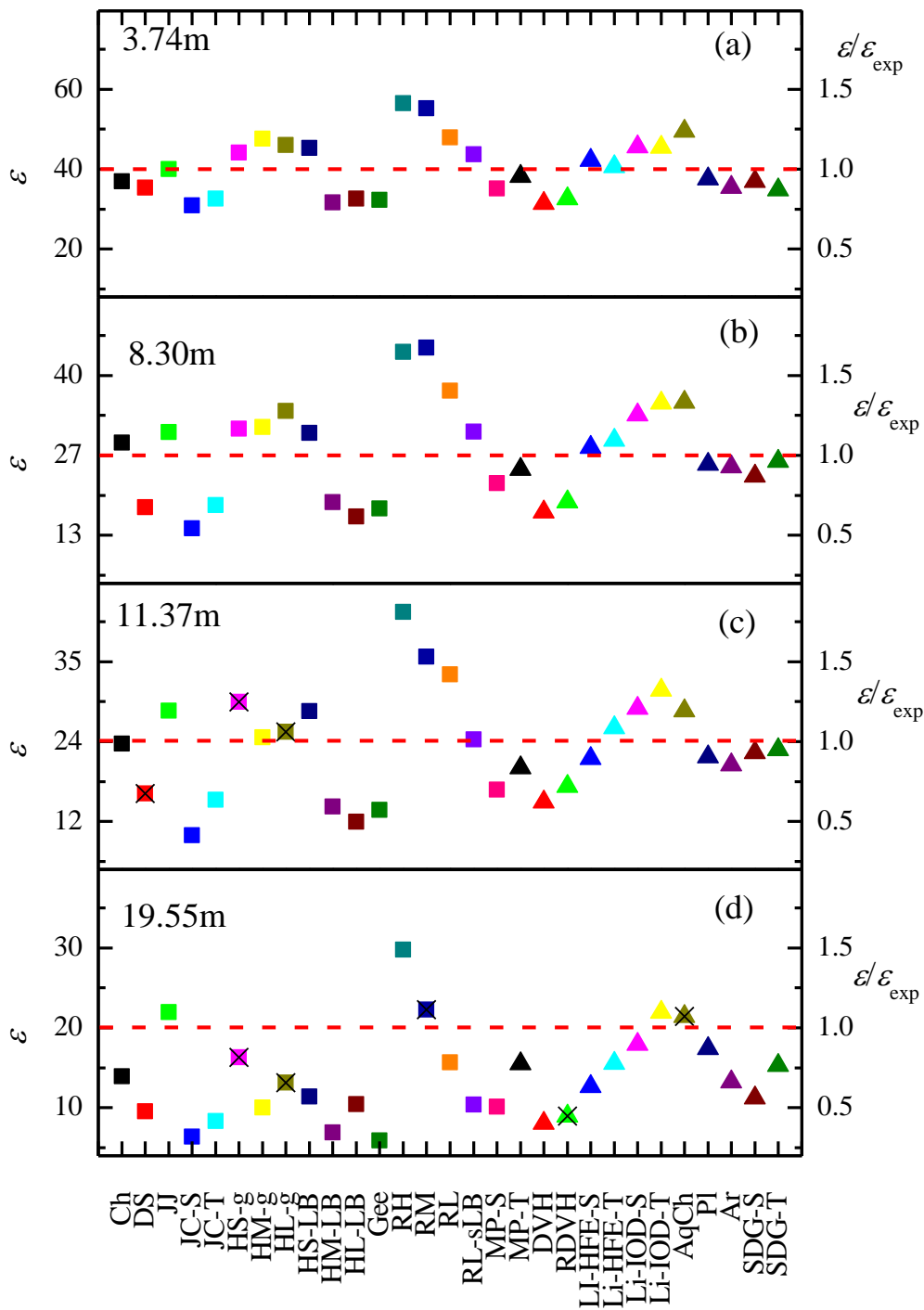

**Fig. 4** Static dielectric constant at the concentrations (a) 3.74 mol/kg, (b) 8.30 mol/kg, (c) 11.37 mol/kg, and (d) 19.55 mol/kg, obtained from the simulations with different FFs. Experimental values are marked with dashed lines. Ratios of the simulated and experimental values ($\varepsilon/\varepsilon_{exp}$) are also shown. Values, which are from definitely not converged $\varepsilon(t)$ curves are crossed out.

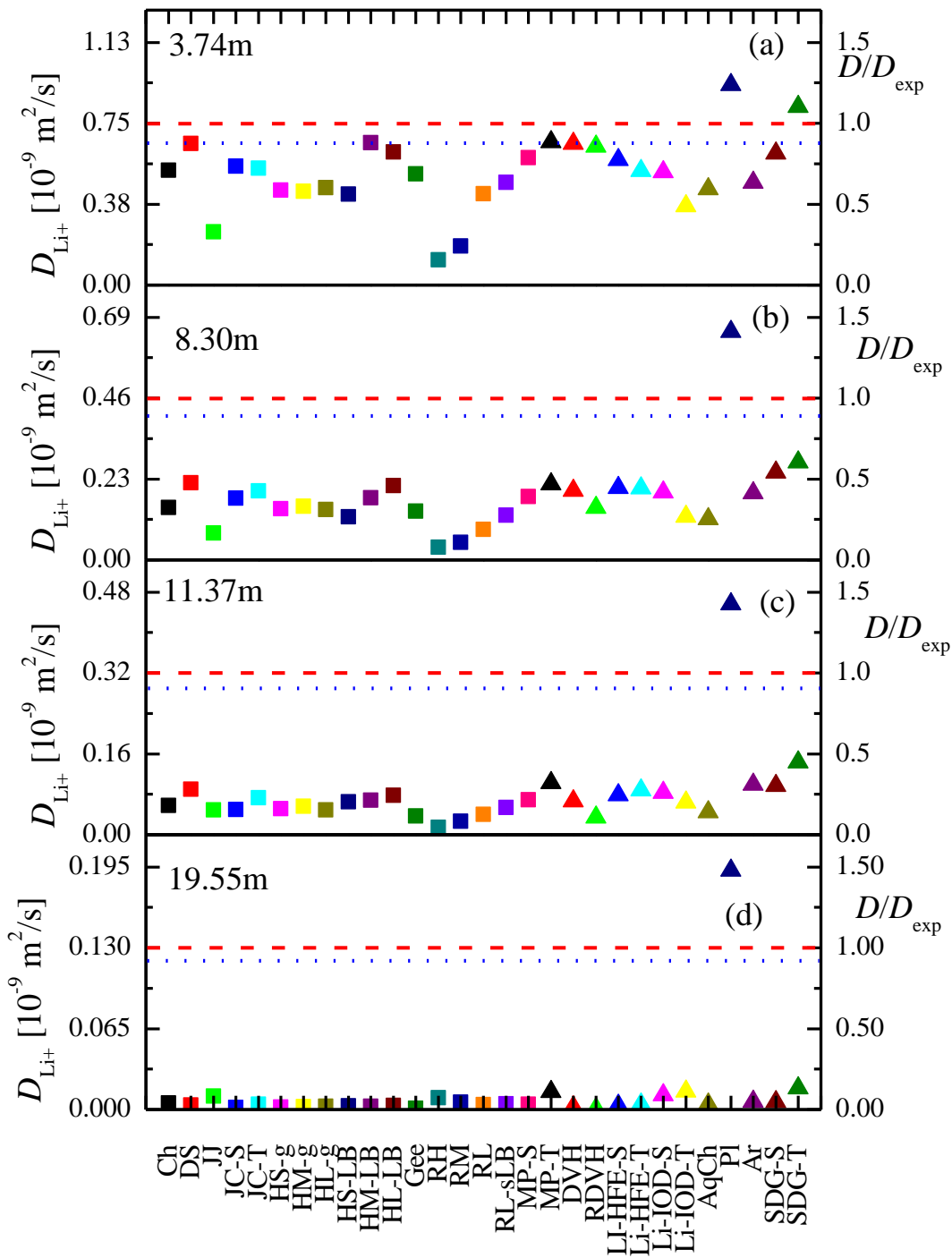

**Fig. 5** Self-diffusion coefficients of the Li$^+$ ions in aqueous LiCl solutions at the concentrations (a) 3.74 mol/kg, (b) 8.30 mol/kg, (c) 11.37 mol/kg, and (d) 19.55 mol/kg, obtained from the simulations with different FFs. The experimental values are marked by dashed lines. The magnitude of the finite size effect is represented as a corrected experimental value by dotted lines.

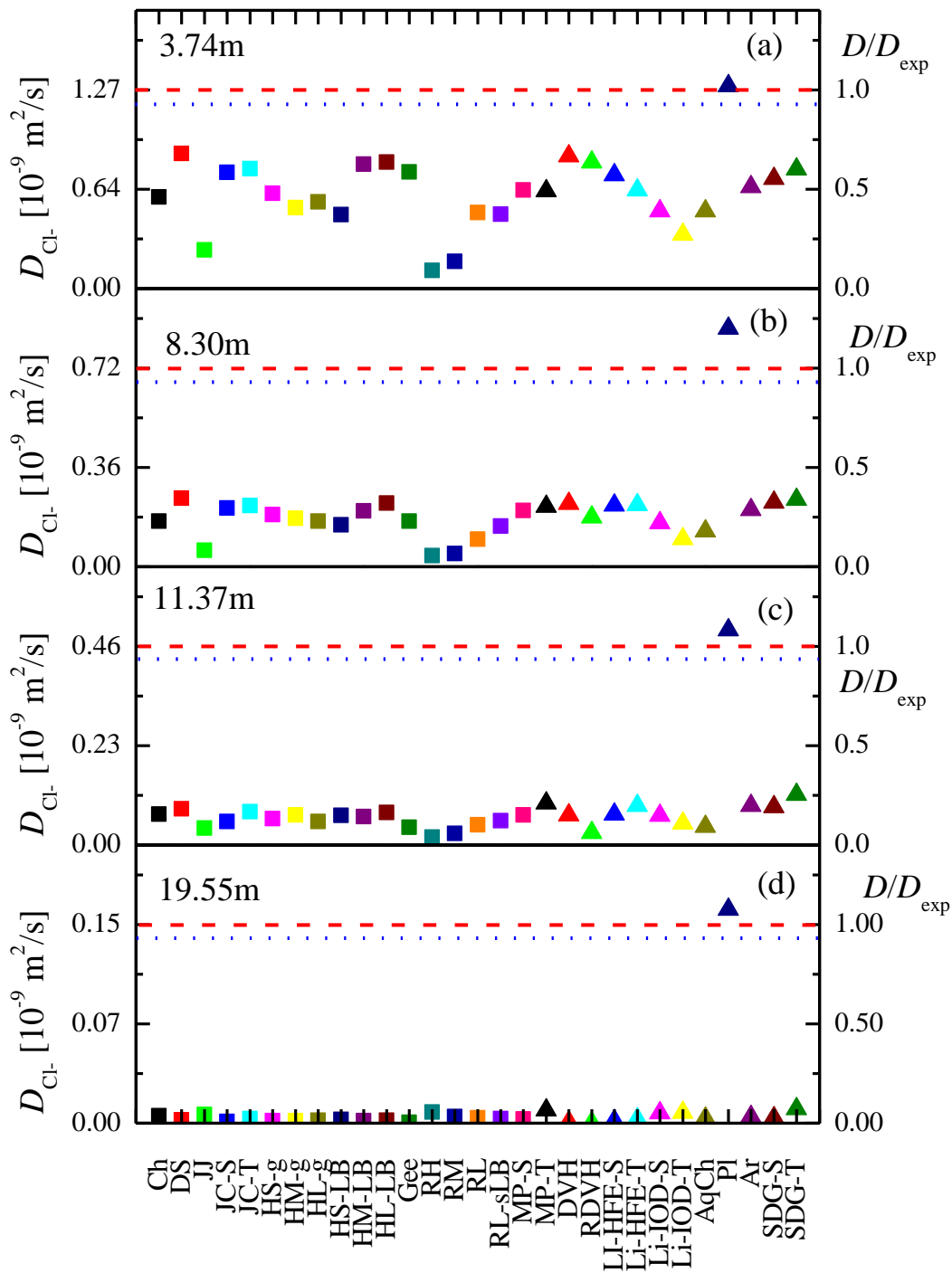

**Fig. 6** Self-diffusion coefficients of the Cl$^-$ ions in aqueous LiCl solutions at the concentrations (a) 3.74 mol/kg, (b) 8.30 mol/kg, (c) 11.37 mol/kg and (d) 19.55 mol/kg, obtained from the simulations with different FFs. The experimental values are marked by dashed lines. The magnitude of the finite size effect is represented as a corrected experimental value by dotted lines.

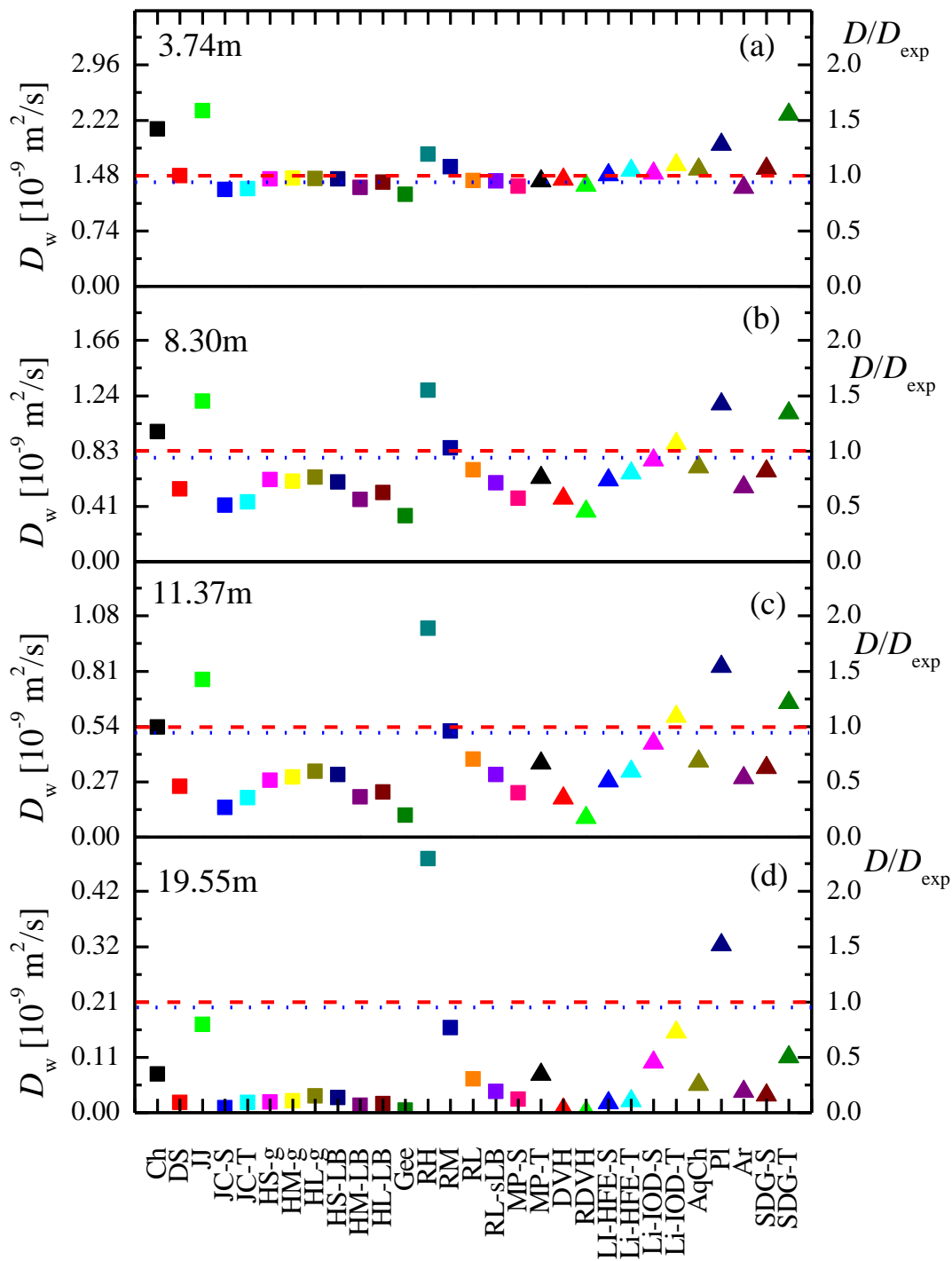

**Fig. 7** Self-diffusion coefficients of water molecules in aqueous LiCl solutions at the concentrations (a) 3.74 mol/kg, (b) 8.30 mol/kg, (c) 11.37 mol/kg, and (d) 19.55 mol/kg, obtained from the simulations with different FFs. The experimental values are marked by dashed lines. The magnitude of the finite size effect is represented as a corrected experimental value and by dotted lines.

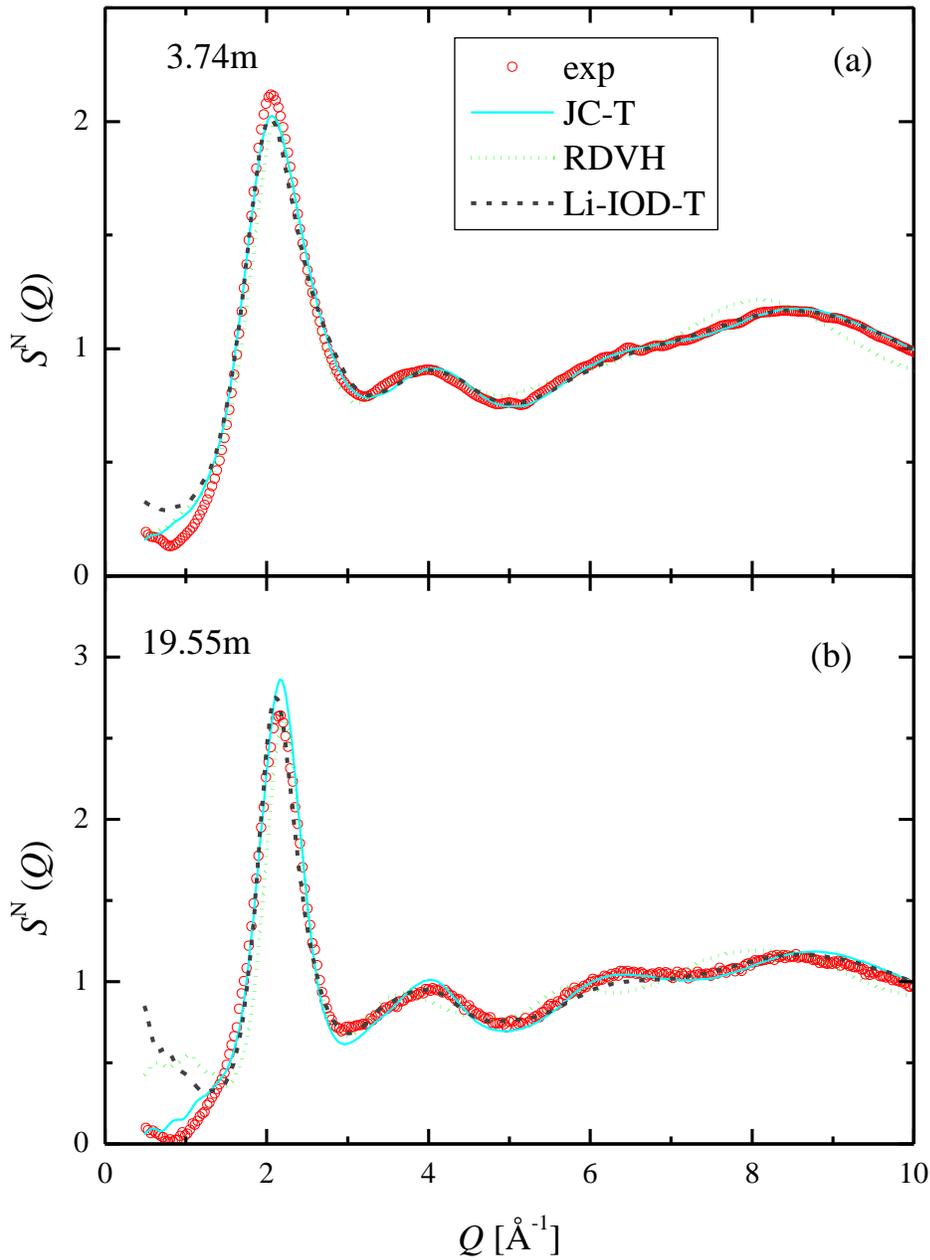

**Fig. 8** Neutron total structure factors from experiments (symbols) and simulations (lines) for the (a) 3.74m sample and (b) the 19.55m sample. The simulated curves were obtained with the FFs JC-T (solid lines), RDVH (dotted lines) and Li-IOD-T (dashed lines).

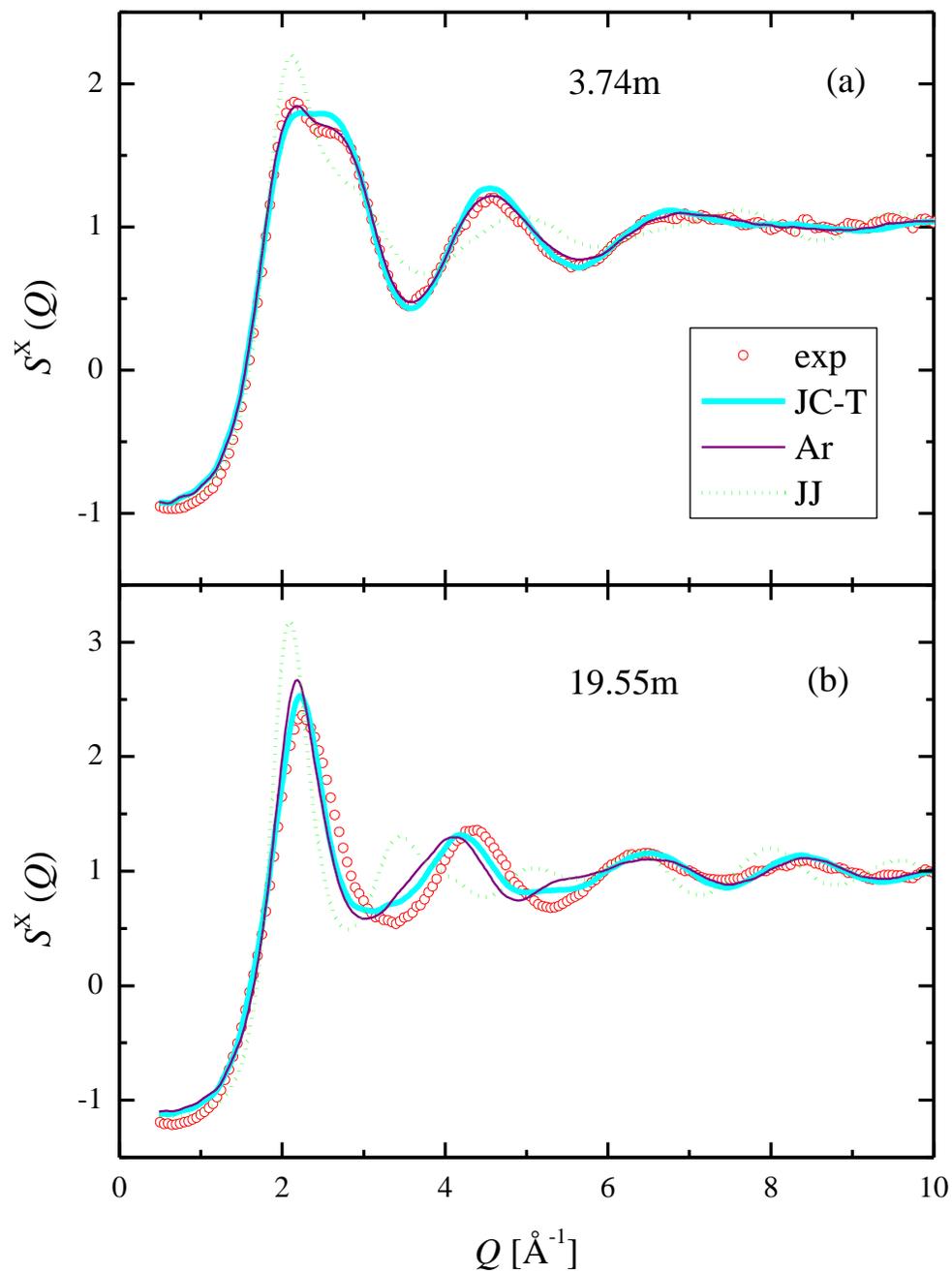

**Fig. 9** X-ray total structure factors from experiments (symbols) and simulations (lines) for the (a) 3.74m sample and (b) the 19.55m sample. The simulated curves were obtained with FFs JC-T (thick solid lines), Ar (thin solid lines) and JJ (dotted lines).

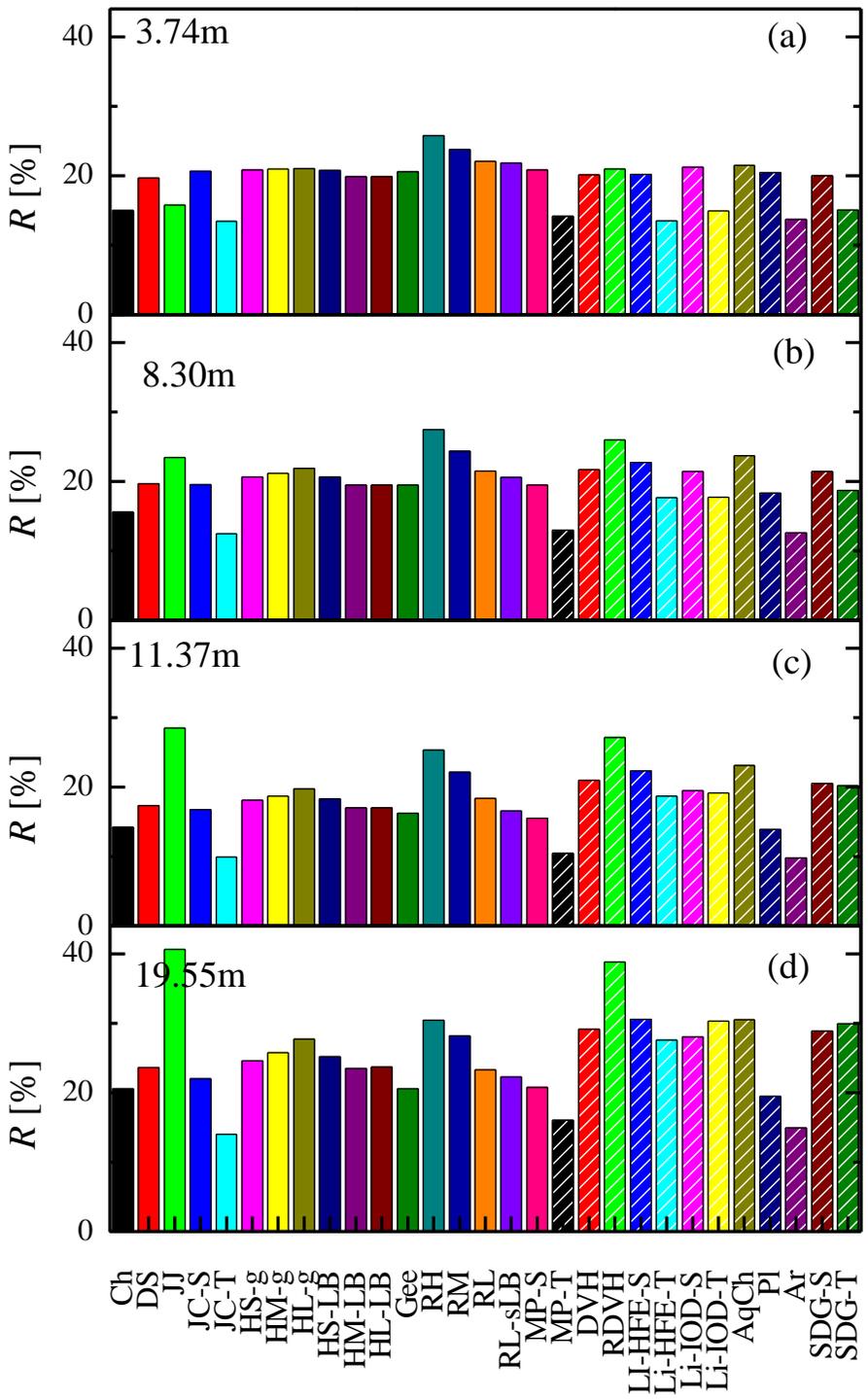

**Fig. 10** *R*-factors (see text) of the simulated neutron structure factors calculated for the different FFs at the concentrations (a) 3.74 mol/kg, (b) 8.30 mol/kg, (c) 11.37 mol/kg, (d) 19.55 mol/kg.

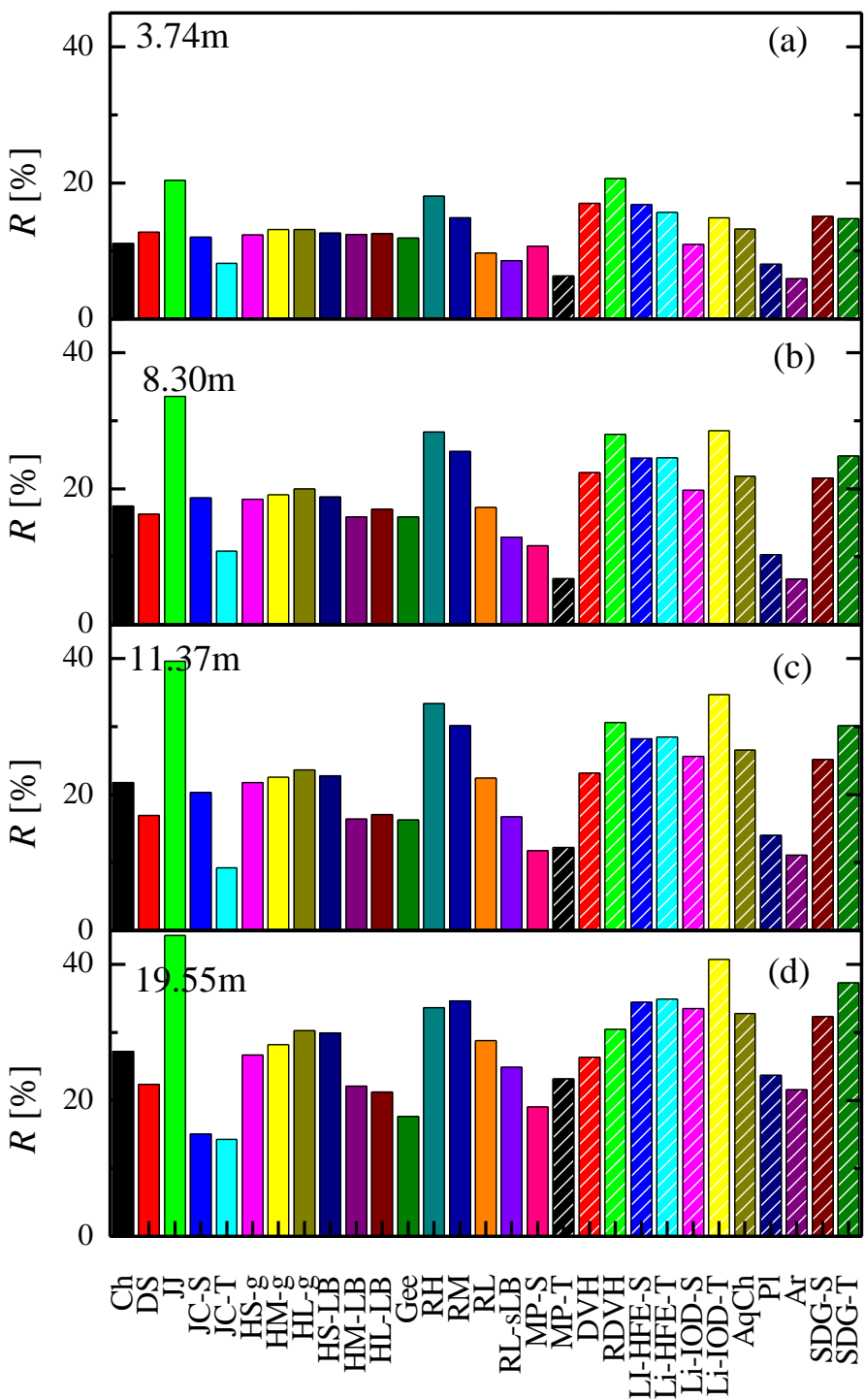

**Fig. 11** *R*-factors (see text) of the simulated X-ray structure factors calculated for the different FFs at the concentrations (a) 3.74 mol/kg, (b) 8.30 mol/kg, (c) 11.37 mol/kg, (d) 19.55 mol/kg.

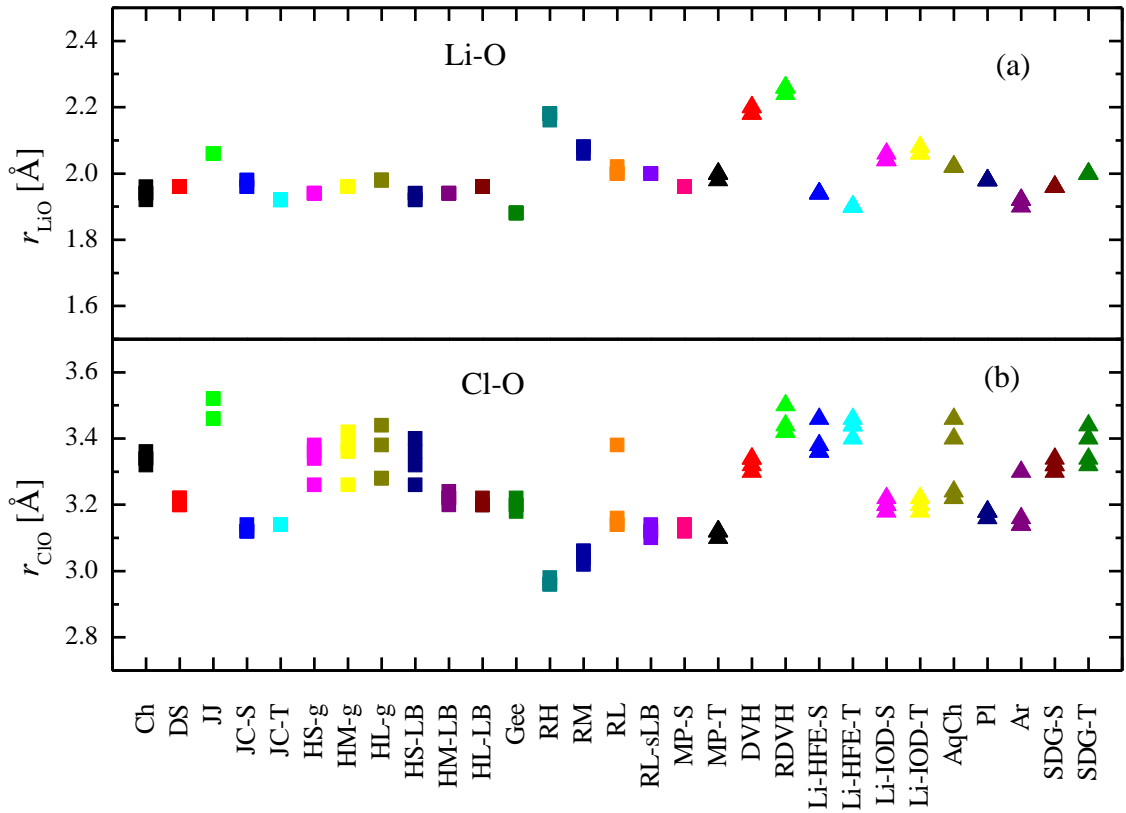

**Fig. 12** (a) Average Li$^+$-O and (b) Cl$^-$-O distances (positions of the firs maximum in the corresponding PPCF) obtained from the simulations with different FFs at four concentrations.

# Supplementary Material

# A comparison of classical interatomic potentials applied to highly concentrated aqueous lithium chloride solutions

Ildikó Pethes

Wigner Research Centre for Physics, Hungarian Academy of Sciences, H-1525 Budapest, POB 49, Hungary

E-mail address: pethes.ildiko@wigner.mta.hu

**Table S.1** Force field parameters of the investigated models. ($\sigma_{ij}$ in nm, $\varepsilon_{ij}$ in kJ/mol)

| | $\sigma_{LiLi}$ | $\varepsilon_{LiLi}$ | $\sigma_{ClCl}$ | $\varepsilon_{ClCl}$ | $\sigma_{LiCl}$ | $\sigma_{LiO}$ | $\sigma_{ClO}$ | $\varepsilon_{LiCl}$ | $\varepsilon_{LiO}$ | $\varepsilon_{ClO}$ | Ref. |
|---|---|---|---|---|---|---|---|---|---|---|---|
| Ch | 0.1260 | 26.1495 | 0.4417 | 0.4928 | 0.2359 | 0.1993 | 0.3732 | 3.5899 | 4.1181 | 0.5653 | 1 |
| DS | 0.1506 | 0.6904 | 0.4400 | 0.4184 | 0.2956 | 0.2337 | 0.3785 | 0.5375 | 0.6700 | 0.5216 | 2,3 |
| JJ | 0.2870 | 0.0021 | 0.4020 | 2.9706 | 0.3397 | 0.3008 | 0.3561 | 0.0788 | 0.0368 | 1.3880 | 4 |
| JC-S | 0.1409 | 1.4089 | 0.4830 | 0.0535 | 0.3120 | 0.2287 | 0.3998 | 0.2742 | 0.9560 | 0.1865 | 5 |
| JC-T | 0.1440 | 0.4351 | 0.4918 | 0.0488 | 0.3179 | 0.2302 | 0.4041 | 0.1457 | 0.5443 | 0.1823 | 5 |
| HS-g | 0.2880 | 0.0006 | 0.4520 | 0.4200 | 0.3608 | 0.3019 | 0.3783 | 0.0160 | 0.0199 | 0.5226 | 6 |
| HM-g | 0.1700 | 0.6500 | 0.4520 | 0.4200 | 0.2772 | 0.2320 | 0.3783 | 0.5225 | 0.6501 | 0.5226 | 6 |
| HL-g | 0.1630 | 1.5400 | 0.4520 | 0.4200 | 0.2714 | 0.2272 | 0.3783 | 0.8042 | 1.0006 | 0.5226 | 6 |
| HS-LB | 0.2870 | 0.0006 | 0.4400 | 0.4200 | 0.3635 | 0.3018 | 0.3783 | 0.0160 | 0.0199 | 0.5226 | 6 |
| HM-LB | 0.1470 | 0.6500 | 0.4400 | 0.4200 | 0.2935 | 0.2318 | 0.3783 | 0.5225 | 0.6501 | 0.5226 | 6 |
| HL-LB | 0.1370 | 1.5400 | 0.4400 | 0.4200 | 0.2885 | 0.2268 | 0.3783 | 0.8042 | 1.0006 | 0.5226 | 6 |
| Gee | 0.182 | 0.7 | 0.44 | 0.47 | 0.2830 | 0.2400 | 0.3732 | 0.5736 | 0.2699 | 0.5528 | 7 |
| RH | 0.3529 | 0.0007 | 0.3493 | 1.7625 | 0.3511 | 0.3342 | 0.3325 | 0.0342 | 0.0207 | 1.0705 | 8 |
| RM | 0.3078 | 0.0015 | 0.3771 | 1.1137 | 0.3407 | 0.3121 | 0.3455 | 0.0409 | 0.0313 | 0.8510 | 8 |
| RL | 0.2679 | 0.0035 | 0.4096 | 0.6785 | 0.3313 | 0.2912 | 0.3601 | 0.0484 | 0.0474 | 0.6642 | 8 |
| RL-sLB | 0.2679 | 0.0035 | 0.4096 | 0.6785 | 0.3388 | 0.2912 | 0.3601 | 0.0484 | 0.0474 | 0.6642 | 8 |
| MP-S | 0.1715 | 0.2412 | 0.4612 | 0.1047 | 0.3164 | 0.2440 | 0.3889 | 0.1589 | 0.3961 | 0.2609 | 9 |
| MP-T | 0.1715 | 0.2412 | 0.4612 | 0.1047 | 0.3164 | 0.2440 | 0.3889 | 0.1589 | 0.4053 | 0.2670 | 9 |
| DVH | 0.1880 | 0.8314 | 0.4410 | 0.8314 | 0.3145 | 0.2523 | 0.3788 | 0.8314 | 0.7353 | 0.7353 | 10 |
| RDVH | 0.1880 | 1.6629 | 0.4410 | 1.6629 | 0.3145 | 0.2523 | 0.3788 | 1.6629 | 1.0398 | 1.0398 | 11 |
| Li-HFE-S | 0.2242 | 0.0115 | 0.4112 | 2.6931 | 0.3177 | 0.2704 | 0.3639 | 0.1757 | 0.0864 | 1.3233 | 12 |
| Li-HFE-T | 0.2184 | 0.0071 | 0.4136 | 2.7309 | 0.3160 | 0.2674 | 0.3650 | 0.1388 | 0.0693 | 1.3637 | 12 |
| Li-IOD-S | 0.2343 | 0.0249 | 0.3852 | 2.2240 | 0.3098 | 0.2754 | 0.3509 | 0.2353 | 0.1272 | 1.2025 | 12 |
| Li-IOD-T | 0.2343 | 0.0249 | 0.3852 | 2.2240 | 0.3098 | 0.2754 | 0.3508 | 0.2353 | 0.1302 | 1.2306 | 12 |
| AqCh | 0.2126 | 0.0765 | 0.4417 | 0.4928 | 0.3065 | 0.2594 | 0.3739 | 0.1941 | 0.2230 | 0.5661 | 1,13 |
| Pl | 0.1800 | 0.0765 | 0.4100 | 0.4928 | 0.2950 | 0.2483 | 0.3633 | 0.1941 | 0.2230 | 0.5661 | 14,15 |
| Ar | 0.1440 | 0.4351 | 0.4918 | 0.0488 | 0.2963 | 0.2302 | 0.4041 | 0.2739 | 0.5443 | 0.1823 | 16 |
| SDG-S | 0.1506 | 0.6945 | 0.402 | 2.9706 | 0.2763 | 0.2336 | 0.3593 | 1.4364 | 0.6720 | 1.3898 | 17 |
| SDG-T | 0.1506 | 0.6945 | 0.402 | 2.9706 | 0.2763 | 0.2330 | 0.3587 | 1.4364 | 0.6711 | 1.388 | 17 |

**Table S.2** Densities of the samples obtained in MD simulations with different force fields (in g/cm$^3$).

|         | 3.74m  | 8.30m  | 11.37m | 19.55m |
|---------|--------|--------|--------|--------|
| Exp (Ref. 18) | 1.0763 | 1.1510 | 1.1950 | 1.2862 |
| Ch      | 1.0703 | 1.1419 | 1.1814 | 1.2633 |
| DS      | 1.0493 | 1.0922 | 1.1147 | 1.1627 |
| JJ      | 1.0608 | 1.1193 | 1.1509 | 1.2105 |
| JC-S    | 1.0680 | 1.1201 | 1.1450 | 1.1948 |
| JC-T    | 1.0605 | 1.1167 | 1.1477 | 1.2126 |
| HS-g    | 1.0535 | 1.1071 | 1.1373 | 1.2008 |
| HM-g    | 1.0530 | 1.1057 | 1.1346 | 1.1989 |
| HL-g    | 1.0513 | 1.1014 | 1.1299 | 1.1917 |
| HS-LB   | 1.0535 | 1.1079 | 1.1395 | 1.2069 |
| HM-LB   | 1.0527 | 1.0967 | 1.1218 | 1.1752 |
| HL-LB   | 1.0557 | 1.0972 | 1.1176 | 1.1634 |
| Gee     | 1.0669 | 1.1277 | 1.1609 | 1.2287 |
| RH      | 1.1088 | 1.2072 | 1.2796 | 1.4052 |
| RM      | 1.0886 | 1.1707 | 1.2176 | 1.2836 |
| RL      | 1.0760 | 1.1506 | 1.1933 | 1.2803 |
| RL-sLB  | 1.0770 | 1.1477 | 1.1858 | 1.2658 |
| MP-S    | 1.0670 | 1.1233 | 1.1526 | 1.2136 |
| MP-T    | 1.0781 | 1.1334 | 1.1603 | 1.2188 |
| DVH     | 1.0662 | 1.1047 | 1.1139 | 1.1177 |
| RDVH    | 1.0534 | 1.0861 | 1.0997 | 1.1244 |
| Li-HFE-S | 1.0493 | 1.0939 | 1.1171 | 1.1691 |
| Li-HFE-T | 1.0553 | 1.1069 | 1.1332 | 1.1873 |
| Li-IOD-S | 1.0796 | 1.1447 | 1.1783 | 1.2458 |
| Li-IOD-T | 1.0819 | 1.1505 | 1.1868 | 1.2570 |
| AqCh    | 1.0520 | 1.1027 | 1.1317 | 1.1918 |
| Pl      | 1.0395 | 1.0842 | 1.1115 | 1.1720 |
| Ar      | 1.0590 | 1.1166 | 1.1499 | 1.2191 |
| SDG-S   | 1.0612 | 1.1123 | 1.1376 | 1.1901 |
| SDG-T   | 1.0774 | 1.1361 | 1.1635 | 1.2121 |

Table S.3 Static dielectric constants of the samples obtained in MD simulations with different force fields.

|          | 3.74m | 8.30m | 11.37m | 19.55m |
|----------|-------|-------|--------|--------|
| Exp (Ref.19) | 40.1 | 26.8 | 23.5 | 19.9 |
| Ch       | 37.0 | 28.9 | 23.2 | 13.8 |
| DS       | 35.5 | 18.1 | 15.9 | 9.5 |
| JJ       | 40.1 | 30.7 | 28.0 | 21.8 |
| JC-S     | 31.0 | 14.6 | 9.8 | 6.4 |
| JC-T     | 32.7 | 18.4 | 15.0 | 8.3 |
| HS-g     | 44.3 | 31.3 | 29.3 | 16.2 |
| HM-g     | 47.7 | 31.6 | 24.1 | 10.0 |
| HL-g     | 46.2 | 34.2 | 24.9 | 13.1 |
| HS-LB    | 45.4 | 30.5 | 28.0 | 11.3 |
| HM-LB    | 31.8 | 19.0 | 14.0 | 6.9 |
| HL-LB    | 32.7 | 16.6 | 11.7 | 10.4 |
| Gee      | 32.3 | 17.9 | 13.5 | 5.8 |
| RH       | 56.7 | 44.1 | 42.5 | 29.6 |
| RM       | 55.3 | 44.8 | 36.0 | 22.2 |
| RL       | 48.1 | 37.6 | 33.3 | 15.6 |
| RL-sLB   | 43.8 | 30.8 | 23.8 | 10.3 |
| MP-S     | 35.2 | 22.1 | 16.4 | 10.1 |
| MP-T     | 38.3 | 24.4 | 19.6 | 15.5 |
| DVH      | 31.5 | 17.3 | 14.6 | 8.0 |
| RDVH     | 32.6 | 19.0 | 16.9 | 8.9 |
| Li-HFE-S | 42.4 | 28.2 | 21.0 | 12.6 |
| Li-HFE-T | 40.8 | 29.3 | 25.5 | 15.5 |
| Li-IOD-S | 45.8 | 33.6 | 28.4 | 17.8 |
| Li-IOD-T | 45.6 | 35.6 | 31.1 | 21.8 |
| AqCh     | 49.6 | 35.7 | 28.0 | 21.3 |
| Pl       | 37.7 | 25.3 | 21.3 | 17.3 |
| Ar       | 35.5 | 24.9 | 20.1 | 13.2 |
| SDG-S    | 37.0 | 23.4 | 21.9 | 11.2 |
| SDG-T    | 34.9 | 25.8 | 22.3 | 15.2 |

Table S.4 Li$^+$ ion self-diffusion coefficients of the samples obtained in MD simulations with different force fields (in 10$^{-9}$ m$^2$/s).

|           | 3.74m | 8.30m | 11.37m | 19.55m |
|-----------|-------|-------|--------|--------|
| Exp (Ref. 20) | 0.75 | 0.46 | 0.32 | 0.13 |
| Ch        | 0.533 | 0.149 | 0.058 | 0.005 |
| DS        | 0.657 | 0.220 | 0.090 | 0.004 |
| JJ        | 0.247 | 0.076 | 0.049 | 0.011 |
| JC-S      | 0.550 | 0.175 | 0.050 | 0.002 |
| JC-T      | 0.542 | 0.196 | 0.073 | 0.004 |
| HS-g      | 0.440 | 0.145 | 0.052 | 0.002 |
| HM-g      | 0.434 | 0.152 | 0.056 | 0.002 |
| HL-g      | 0.452 | 0.143 | 0.049 | 0.003 |
| HS-LB     | 0.421 | 0.122 | 0.065 | 0.003 |
| HM-LB     | 0.661 | 0.176 | 0.068 | 0.002 |
| HL-LB     | 0.618 | 0.211 | 0.078 | 0.003 |
| Gee       | 0.515 | 0.138 | 0.037 | 0.001 |
| RH        | 0.117 | 0.036 | 0.015 | 0.009 |
| RM        | 0.181 | 0.050 | 0.027 | 0.006 |
| RL        | 0.424 | 0.087 | 0.040 | 0.004 |
| RL-sLB    | 0.476 | 0.126 | 0.054 | 0.004 |
| MP-S      | 0.590 | 0.180 | 0.069 | 0.004 |
| MP-T      | 0.665 | 0.216 | 0.103 | 0.015 |
| DVH       | 0.657 | 0.198 | 0.067 | 0.002 |
| RDVH      | 0.643 | 0.149 | 0.035 | 0.001 |
| Li-HFE-S  | 0.581 | 0.205 | 0.079 | 0.003 |
| Li-HFE-T  | 0.531 | 0.204 | 0.088 | 0.004 |
| Li-IOD-S  | 0.524 | 0.192 | 0.084 | 0.012 |
| Li-IOD-T  | 0.367 | 0.123 | 0.065 | 0.015 |
| AqCh      | 0.446 | 0.117 | 0.045 | 0.004 |
| Pl        | 0.930 | 0.650 | 0.457 | 0.192 |
| Ar        | 0.475 | 0.190 | 0.100 | 0.007 |
| SDG-S     | 0.611 | 0.250 | 0.097 | 0.006 |
| SDG-T     | 0.828 | 0.278 | 0.143 | 0.017 |

Table S.5 Cl⁻ ion self-diffusion coefficients of the samples obtained in MD simulations with different force fields (in $10^{-9}$ m$^2$/s).

|  | 3.74m | 8.3m | 11.37m | 19.55m |
|---|---|---|---|---|
| Exp (Ref. 20) | 1.27 | 0.72 | 0.46 | 0.15 |
| Ch | 0.585 | 0.164 | 0.071 | 0.005 |
| DS | 0.864 | 0.249 | 0.084 | 0.002 |
| JJ | 0.246 | 0.060 | 0.039 | 0.007 |
| JC-S | 0.742 | 0.213 | 0.054 | 0.001 |
| JC-T | 0.767 | 0.222 | 0.077 | 0.003 |
| HS-g | 0.609 | 0.188 | 0.061 | 0.002 |
| HM-g | 0.517 | 0.175 | 0.069 | 0.002 |
| HL-g | 0.554 | 0.166 | 0.054 | 0.002 |
| HS-LB | 0.472 | 0.152 | 0.068 | 0.003 |
| HM-LB | 0.796 | 0.202 | 0.066 | 0.002 |
| HL-LB | 0.809 | 0.231 | 0.075 | 0.002 |
| Gee | 0.744 | 0.165 | 0.041 | 0.001 |
| RH | 0.116 | 0.040 | 0.018 | 0.008 |
| RM | 0.173 | 0.047 | 0.026 | 0.005 |
| RL | 0.485 | 0.100 | 0.047 | 0.004 |
| RL-sLB | 0.474 | 0.148 | 0.056 | 0.004 |
| MP-S | 0.630 | 0.203 | 0.069 | 0.003 |
| MP-T | 0.624 | 0.218 | 0.097 | 0.010 |
| DVH | 0.848 | 0.229 | 0.070 | 0.001 |
| RDVH | 0.809 | 0.180 | 0.029 | 0.0004 |
| Li-HFE-S | 0.726 | 0.223 | 0.072 | 0.002 |
| Li-HFE-T | 0.630 | 0.224 | 0.092 | 0.003 |
| Li-IOD-S | 0.497 | 0.159 | 0.069 | 0.008 |
| Li-IOD-T | 0.345 | 0.102 | 0.051 | 0.008 |
| AqCh | 0.495 | 0.129 | 0.043 | 0.004 |
| Pl | 1.296 | 0.864 | 0.497 | 0.161 |
| Ar | 0.651 | 0.207 | 0.091 | 0.005 |
| SDG-S | 0.703 | 0.234 | 0.088 | 0.004 |
| SDG-T | 0.763 | 0.244 | 0.116 | 0.011 |

Table S.6 H$_2$O self-diffusion coefficients of the samples obtained in MD simulations with different force fields (in 10$^{-9}$ m$^2$/s).

|  | 3.74m | 8.3m | 11.37m | 19.55m |
|---|---|---|---|---|
| Exp (Refs.20,21) | 1.48 | 0.83 | 0.54 | 0.21 |
| Ch | 2.101 | 0.976 | 0.538 | 0.073 |
| DS | 1.477 | 0.545 | 0.247 | 0.019 |
| JJ | 2.346 | 1.204 | 0.770 | 0.168 |
| JC-S | 1.296 | 0.425 | 0.146 | 0.009 |
| JC-T | 1.302 | 0.447 | 0.193 | 0.019 |
| HS-g | 1.433 | 0.615 | 0.277 | 0.020 |
| HM-g | 1.446 | 0.604 | 0.294 | 0.022 |
| HL-g | 1.441 | 0.635 | 0.321 | 0.032 |
| HS-LB | 1.436 | 0.599 | 0.306 | 0.028 |
| HM-LB | 1.322 | 0.468 | 0.197 | 0.014 |
| HL-LB | 1.393 | 0.517 | 0.219 | 0.017 |
| Gee | 1.226 | 0.344 | 0.107 | 0.005 |
| RH | 1.769 | 1.287 | 1.020 | 0.482 |
| RM | 1.601 | 0.855 | 0.518 | 0.161 |
| RL | 1.412 | 0.690 | 0.381 | 0.064 |
| RL-sLB | 1.410 | 0.591 | 0.306 | 0.040 |
| MP-S | 1.337 | 0.477 | 0.215 | 0.025 |
| MP-T | 1.407 | 0.632 | 0.362 | 0.073 |
| DVH | 1.427 | 0.475 | 0.191 | 0.006 |
| RDVH | 1.350 | 0.378 | 0.095 | 0.001 |
| Li-HFE-S | 1.490 | 0.613 | 0.274 | 0.018 |
| Li-HFE-T | 1.547 | 0.666 | 0.322 | 0.022 |
| Li-IOD-S | 1.509 | 0.761 | 0.458 | 0.096 |
| Li-IOD-T | 1.624 | 0.886 | 0.589 | 0.153 |
| AqCh | 1.566 | 0.711 | 0.371 | 0.053 |
| Pl | 1.897 | 1.183 | 0.832 | 0.318 |
| Ar | 1.323 | 0.560 | 0.290 | 0.040 |
| SDG-S | 1.574 | 0.679 | 0.338 | 0.033 |
| SDG-T | 2.297 | 1.116 | 0.657 | 0.105 |

**Table S.7** $R$-factors of the neutron total scattering factors ($S^N(Q)$) for the four samples obtained in MD simulations with different force fields (in %).

|         | 3.74m | 8.30m | 11.37m | 19.55m |
|---------|-------|-------|--------|--------|
| Ch      | 14.98 | 15.55 | 14.22  | 20.59  |
| DS      | 19.67 | 19.69 | 17.33  | 23.64  |
| JJ      | 15.77 | 23.44 | 28.47  | 40.64  |
| JC-S    | 20.66 | 19.54 | 16.73  | 21.98  |
| JC-T    | 13.44 | 12.46 | 9.94   | 14.04  |
| HS-g    | 20.80 | 20.65 | 18.11  | 24.60  |
| HM-g    | 20.96 | 21.14 | 18.68  | 25.76  |
| HL-g    | 21.06 | 21.86 | 19.73  | 27.69  |
| HS-LB   | 20.76 | 20.65 | 18.31  | 25.16  |
| HM-LB   | 19.87 | 19.49 | 16.99  | 23.47  |
| HL-LB   | 19.84 | 19.45 | 17.00  | 23.72  |
| Gee     | 20.57 | 19.48 | 16.25  | 20.60  |
| RH      | 25.79 | 27.44 | 25.29  | 30.45  |
| RM      | 23.76 | 24.37 | 22.14  | 28.18  |
| RL      | 22.06 | 21.46 | 18.34  | 23.27  |
| RL-sLB  | 21.82 | 20.56 | 16.56  | 22.27  |
| MP-S    | 20.80 | 19.45 | 15.54  | 20.80  |
| MP-T    | 14.18 | 12.97 | 10.43  | 16.01  |
| DVH     | 20.15 | 21.64 | 20.93  | 29.15  |
| RDVH    | 20.98 | 25.94 | 27.13  | 38.81  |
| Li-HFE-S| 20.21 | 22.73 | 22.30  | 30.59  |
| Li-HFE-T| 13.53 | 17.65 | 18.68  | 27.60  |
| Li-IOD-S| 21.22 | 21.40 | 19.45  | 28.03  |
| Li-IOD-T| 14.93 | 17.71 | 19.16  | 30.28  |
| AqCh    | 21.47 | 23.69 | 23.09  | 30.51  |
| Pl      | 20.47 | 18.32 | 13.86  | 19.45  |
| Ar      | 13.72 | 12.59 | 9.83   | 14.94  |
| SDG-S   | 19.98 | 21.39 | 20.50  | 28.89  |
| SDG-T   | 15.05 | 18.67 | 20.17  | 29.92  |

**Table S.8** $R$-factors of the X-ray total scattering factors ($S^X(Q)$) for the four samples obtained in MD simulations with different force fields (in %).

|         | 3.74m | 8.30m | 11.37m | 19.55m |
|---------|-------|-------|--------|--------|
| Ch      | 11.11 | 17.47 | 21.74  | 27.24  |
| DS      | 12.76 | 16.28 | 16.92  | 22.34  |
| JJ      | 20.35 | 33.60 | 39.59  | 44.29  |
| JC-S    | 12.04 | 18.67 | 20.29  | 15.08  |
| JC-T    | 8.14  | 10.83 | 9.20   | 14.26  |
| HS-g    | 12.35 | 18.47 | 21.75  | 26.65  |
| HM-g    | 13.12 | 19.10 | 22.57  | 28.23  |
| HL-g    | 13.14 | 19.95 | 23.61  | 30.28  |
| HS-LB   | 12.60 | 18.81 | 22.76  | 29.90  |
| HM-LB   | 12.38 | 15.87 | 16.37  | 22.09  |
| HL-LB   | 12.56 | 17.02 | 17.05  | 21.25  |
| Gee     | 11.86 | 15.84 | 16.26  | 17.69  |
| RH      | 18.08 | 28.34 | 33.41  | 33.66  |
| RM      | 14.90 | 25.47 | 30.12  | 34.66  |
| RL      | 9.68  | 17.23 | 22.43  | 28.79  |
| RL-sLB  | 8.56  | 12.89 | 16.71  | 24.92  |
| MP-S    | 10.67 | 11.63 | 11.74  | 19.03  |
| MP-T    | 6.31  | 6.79  | 12.18  | 23.16  |
| DVH     | 17.01 | 22.37 | 23.15  | 26.37  |
| RDVH    | 20.67 | 27.99 | 30.59  | 30.48  |
| Li-HFE-S| 16.77 | 24.52 | 28.24  | 34.44  |
| Li-HFE-T| 15.66 | 24.57 | 28.46  | 34.88  |
| Li-IOD-S| 10.93 | 19.76 | 25.62  | 33.55  |
| Li-IOD-T| 14.88 | 28.54 | 34.70  | 40.75  |
| AqCh    | 13.22 | 21.84 | 26.54  | 32.80  |
| Pl      | 8.05  | 10.26 | 14.04  | 23.70  |
| Ar      | 5.93  | 6.69  | 11.07  | 21.58  |
| SDG-S   | 15.06 | 21.57 | 25.13  | 32.34  |
| SDG-T   | 14.73 | 24.83 | 30.14  | 37.30  |

Table S.9 First maxima of the Li-O partial pair correlation function obtained in MD simulations with the different force fields (in Å).

|          | 3.74m | 8.3m | 11.37m | 19.55m |
|----------|-------|------|--------|--------|
| Ch       | 1.94  | 1.92 | 1.94   | 1.96   |
| DS       | 1.96  | 1.96 | 1.96   | 1.96   |
| JJ       | 2.06  | 2.06 | 2.06   | 2.06   |
| JC-S     | 1.98  | 1.96 | 1.98   | 1.96   |
| JC-T     | 1.92  | 1.92 | 1.92   | 1.92   |
| HS-g     | 1.94  | 1.94 | 1.94   | 1.94   |
| HM-g     | 1.96  | 1.96 | 1.96   | 1.96   |
| HL-g     | 1.98  | 1.98 | 1.98   | 1.98   |
| HS-LB    | 1.92  | 1.94 | 1.92   | 1.94   |
| HM-LB    | 1.94  | 1.94 | 1.94   | 1.94   |
| HL-LB    | 1.96  | 1.96 | 1.96   | 1.96   |
| Gee      | 1.88  | 1.88 | 1.88   | 1.88   |
| RH       | 2.18  | 2.18 | 2.18   | 2.16   |
| RM       | 2.08  | 2.06 | 2.08   | 2.08   |
| RL       | 2     | 2    | 2.02   | 2      |
| RL-sLB   | 2     | 2    | 2      | 2      |
| MP-S     | 1.96  | 1.96 | 1.96   | 1.96   |
| MP-T     | 1.98  | 2    | 2      | 2      |
| DVH      | 2.2   | 2.2  | 2.18   | 2.2    |
| RDVH     | 2.24  | 2.26 | 2.26   | 2.24   |
| Li-HFE-S | 1.94  | 1.94 | 1.94   | 1.94   |
| Li-HFE-T | 1.9   | 1.9  | 1.9    | 1.9    |
| Li-IOD-S | 2.06  | 2.04 | 2.04   | 2.04   |
| Li-IOD-T | 2.08  | 2.08 | 2.06   | 2.08   |
| AqCh     | 2.02  | 2.02 | 2.02   | 2.02   |
| Pl       | 1.98  | 1.98 | 1.98   | 1.96   |
| Ar       | 1.92  | 1.9  | 1.92   | 1.92   |
| SDG-S    | 1.96  | 1.96 | 1.96   | 1.96   |
| SDG-T    | 2     | 2    | 2      | 1.98   |

Table S.10 First maxima of the Cl-O partial pair correlation function obtained in MD simulations with the different force fields (in Å).

|  | 3.74m | 8.3m | 11.37m | 19.55m |
| --- | --- | --- | --- | --- |
| Ch | 3.32 | 3.34 | 3.34 | 3.36 |
| DS | 3.22 | 3.2 | 3.2 | 3.22 |
| JJ | 3.52 | 3.46 | 3.52 | 3.46 |
| JC-S | 3.14 | 3.12 | 3.12 | 3.12 |
| JC-T | 3.14 | 3.14 | 3.14 | 3.14 |
| HS-g | 3.26 | 3.34 | 3.36 | 3.38 |
| HM-g | 3.26 | 3.36 | 3.38 | 3.42 |
| HL-g | 3.28 | 3.28 | 3.38 | 3.44 |
| HS-LB | 3.26 | 3.32 | 3.36 | 3.4 |
| HM-LB | 3.24 | 3.22 | 3.22 | 3.2 |
| HL-LB | 3.2 | 3.2 | 3.2 | 3.22 |
| Gee | 3.18 | 3.2 | 3.22 | 3.2 |
| RH | 2.98 | 2.96 | 2.96 | 2.96 |
| RM | 3.04 | 3.02 | 3.06 | 3.06 |
| RL | 3.14 | 3.14 | 3.16 | 3.38 |
| RL-sLB | 3.12 | 3.1 | 3.14 | 3.12 |
| MP-S | 3.14 | 3.12 | 3.12 | 3.14 |
| MP-T | 3.12 | 3.1 | 3.12 | 3.12 |
| DVH | 3.32 | 3.3 | 3.34 | 3.34 |
| RDVH | 3.42 | 3.42 | 3.44 | 3.5 |
| Li-HFE-S | 3.36 | 3.36 | 3.38 | 3.46 |
| Li-HFE-T | 3.4 | 3.44 | 3.44 | 3.46 |
| Li-IOD-S | 3.2 | 3.18 | 3.22 | 3.2 |
| Li-IOD-T | 3.18 | 3.22 | 3.2 | 3.22 |
| AqCh | 3.22 | 3.24 | 3.4 | 3.46 |
| Pl | 3.18 | 3.16 | 3.18 | 3.18 |
| Ar | 3.16 | 3.14 | 3.14 | 3.3 |
| SDG-S | 3.3 | 3.32 | 3.32 | 3.34 |
| SDG-T | 3.32 | 3.34 | 3.4 | 3.44 |

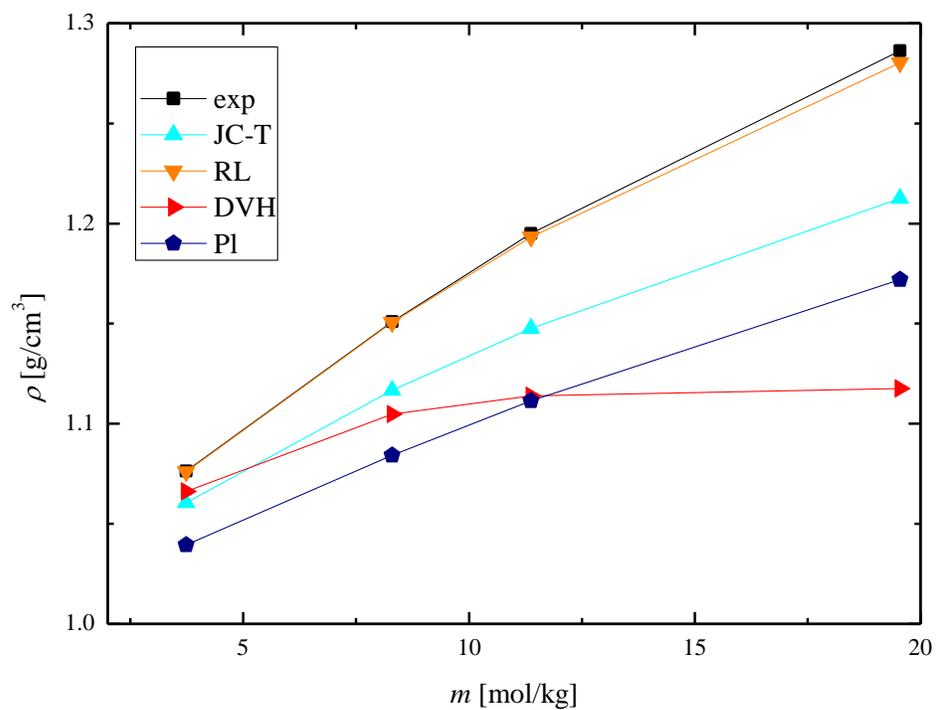

**Figure S.1.** Concentration dependence of the density obtained in simulations with different FFs. Experimental values are also shown.

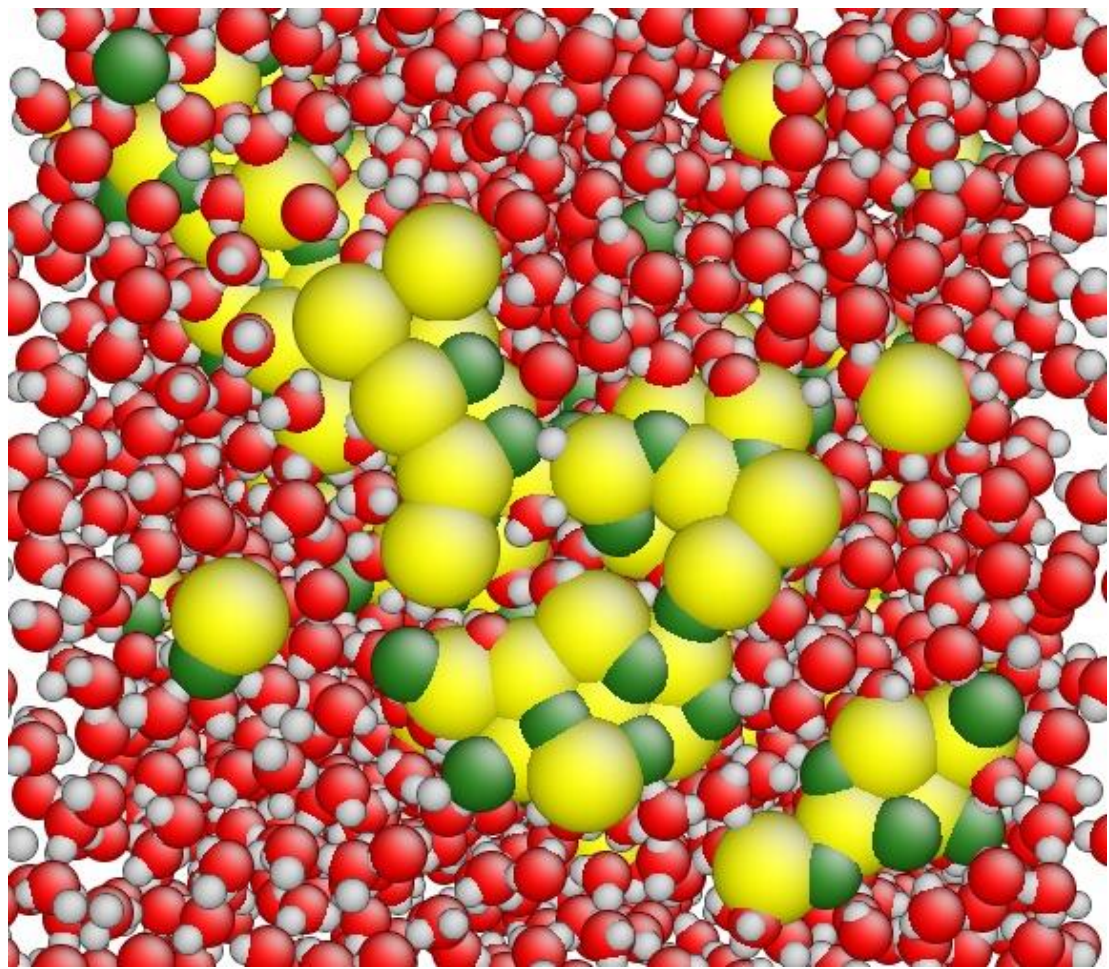

**Figure S.2**. Snapshot of a representative configuration with LiCl precipitate, obtained with the RH FF at the concentration $m$=8.30 mol/kg. Red, gray, yellow and green balls represent oxygen, hydrogen, lithium and chlorine atoms, respectively.

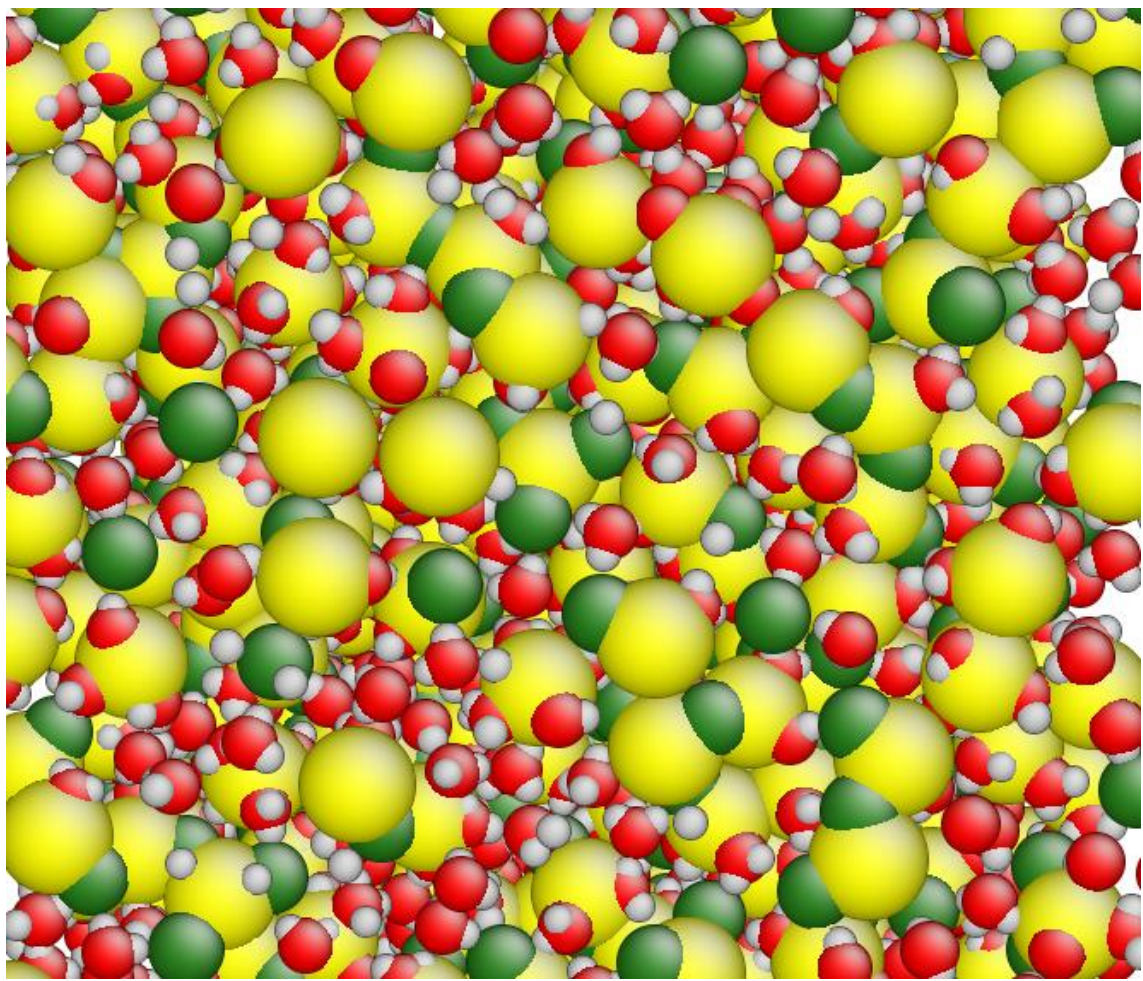

**Figure S.3.** Snapshot of a typical configuration at the concentration *m*=19.55 mol/kg. This configuration was obtained with the MP-T FF. Red, gray, yellow and green balls represent oxygen, hydrogen, lithium and chlorine atoms, respectively.

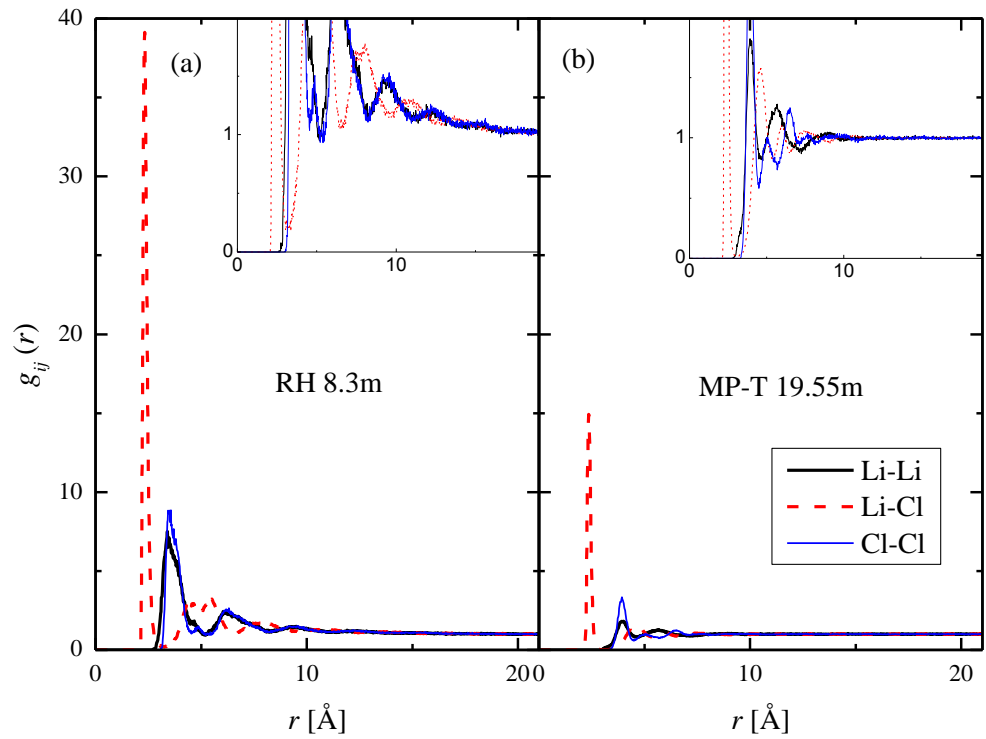

Figure S.4. Ion-ion partial pair correlation functions (a) of the 8.30m sample obtained with the RH FF and (b) of the 19.55m sample obtained with the MP-T FF. The insets highlight the curves around 1: the ion-ion PPCFs for the RH FF slowly converge from above to 1, while for the MP-T FF they oscillate around 1.

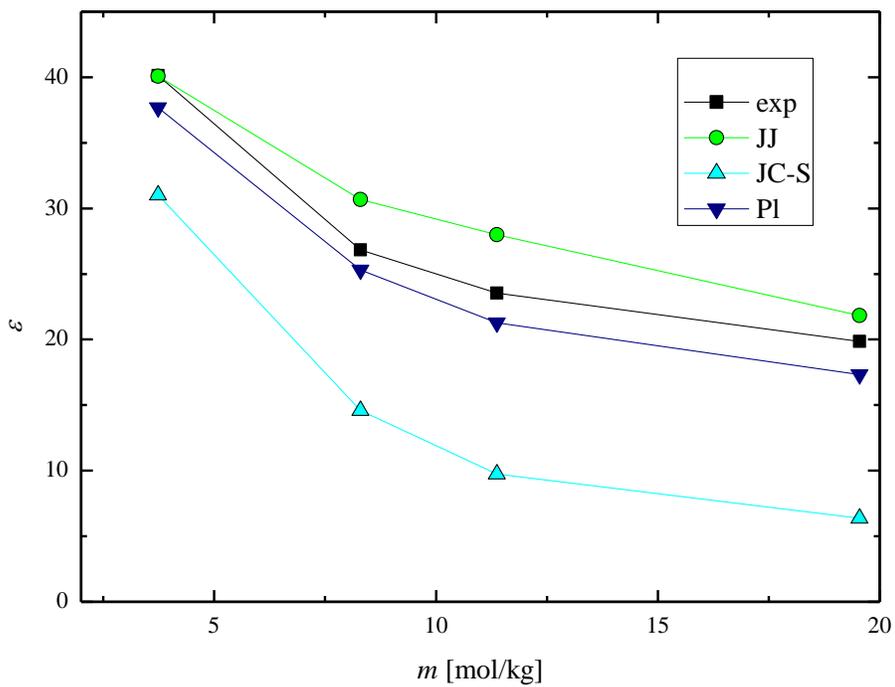

**Figure S.5.** Concentration dependence of the static dielectric constant obtained in simulations with different FFs. Experimental values are also shown.

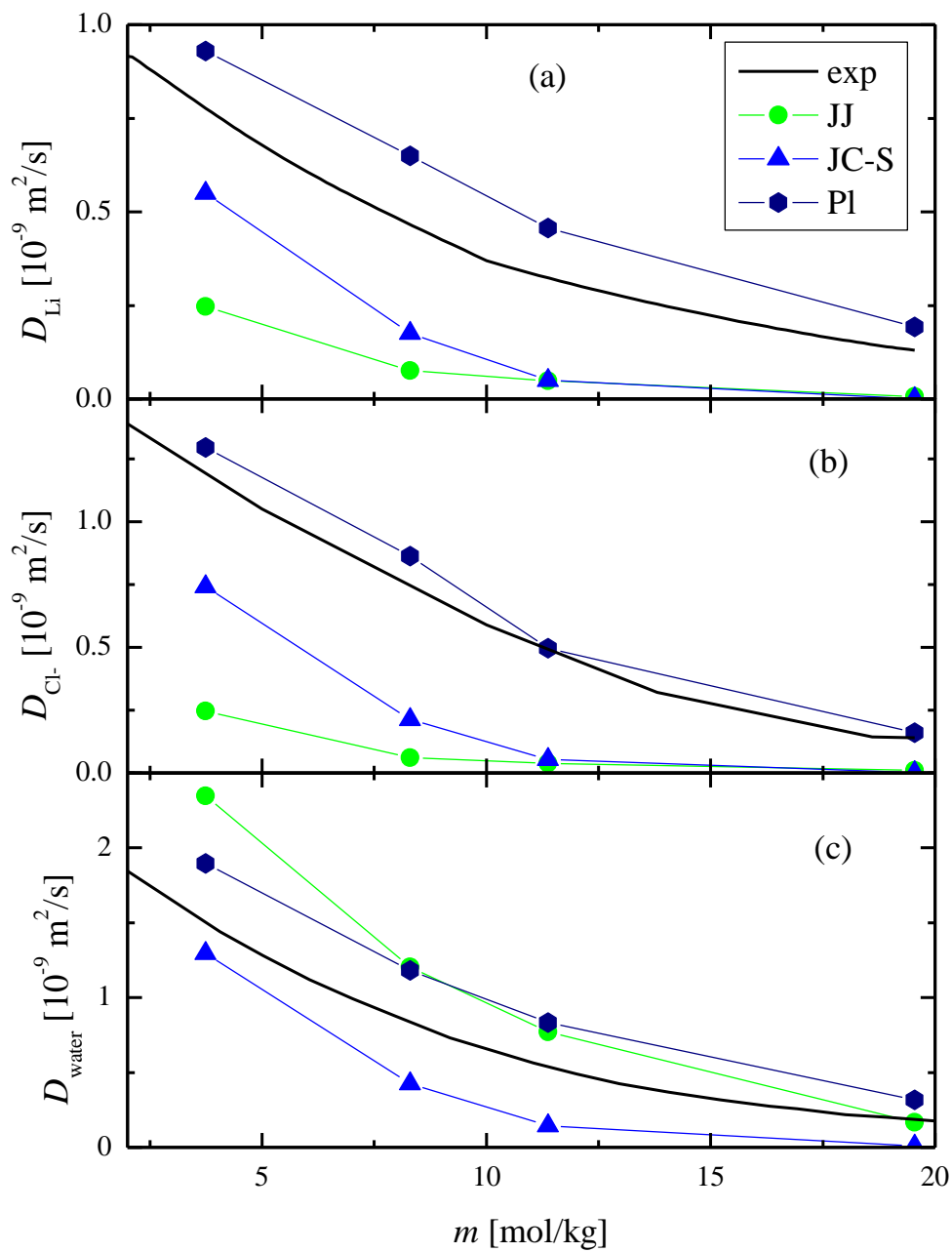

**Figure S.6.** Self-diffusion coefficients of (a) Li$^+$ ions, (b) Cl$^-$ ions and (c) water molecules as a function of concentration, obtained in simulations with different FFs. Experimental values are also shown.